\documentclass[aps,prx,reprint,superscriptaddress,nofootinbib]{revtex4-2}
\usepackage{amsmath,amsthm}
\usepackage{amssymb}
\usepackage{verbatim}
\usepackage{mathtools}
\usepackage{dsfont}
\usepackage{xcolor}
\usepackage{algorithm2e}
\usepackage{graphicx}
\usepackage{subcaption}

\usepackage{nicefrac}

\usepackage{thmtools}
\usepackage{thm-restate}

\usepackage{mathrsfs}
\usepackage{nccmath}
\usepackage{placeins}
\usepackage{physics}
\usepackage{comment}
\usepackage{qcircuit}
\usepackage{pgffor}
\usepackage[colorlinks=true,linkcolor=blue,urlcolor=blue,citecolor=blue]{hyperref}
\usepackage{cleveref}
\usepackage{todonotes}

\newcommand{\bea}{\begin{eqnarray}}
\newcommand{\eea}{\end{eqnarray}}

\newcommand{\DDD}{\color{magenta}}

\newcommand{\Roi}{\color{teal}}

\graphicspath{ {figures} }

\usepackage{ragged2e}

\usepackage[compatibility=false,%
            format=plain,%
            justification=justified,%
            singlelinecheck=false]{caption}

\makeatletter
\renewcommand\@makecaption[2]{%
  \vskip\abovecaptionskip
  {\small #1.\ }\justifying #2\par
  \vskip\belowcaptionskip}
\makeatother

\usepackage{float}

\begin{document}

\title{ 
Quantum Transport in Disordered Spin Networks: Emergent Timescales and Competing Pathways 
}


%


\author{Roi Nevo}
\affiliation{The Racah Institute of Physics, The Hebrew University of Jerusalem, Jerusalem 91904, Israel}
\affiliation{Inst. of Applied Physics, The Hebrew University of Jerusalem, Jerusalem 91904, Israel.}
\author{Brett Min}
\affiliation{Department of Physics, 60 Saint George St., University of Toronto, Toronto, Ontario,  M5S 1A7, Canada}
\author{Maggie Lawrence}
\affiliation{Department of Physics, 60 Saint George St., University of Toronto, Toronto, Ontario,  M5S 1A7, Canada}
\affiliation{Vector Institute, W1140-108 College Street, Schwartz Reisman Innovation Campus Toronto, Ontario M5G 0C6, Canada}

\author{Dvira Segal}
\email{Corresponding author: dvira.segal@utoronto.ca}
\affiliation{Department of Chemistry and Centre for Quantum Information and Quantum Control,
University of Toronto, 80 Saint George St., Toronto, Ontario, M5S 3H6, Canada}
\affiliation{Department of Physics, 60 Saint George St., University of Toronto, Toronto, Ontario,  M5S 1A7, Canada}

\author{Nir Bar-Gill}
\email{Corresponding author: nir.bar-gill@mail.huji.ac.il}
\affiliation{Inst. of Applied Physics, The Hebrew University of Jerusalem, Jerusalem 91904, Israel.}
\affiliation{The Racah Institute of Physics, The Hebrew University of Jerusalem, Jerusalem 91904, Israel}

\begin{abstract}
Quantum transport in disordered systems poses intriguing fundamental questions, such as the interplay of disorder, interactions, and decoherence, and has attracted significant interest. These fundamental aspects also have important implications in real-world scenarios, such as energy transfer in nanoscale systems (e.g. light harvesting complexes)  and quantum information transfer. This large body of work spanning broad research areas in Physics, Chemistry and Nanotechnology has led to major insights and advances, yet open questions remain, namely in terms of the emergence of multiple timescales observed in the dynamics of such systems. 
Here, we investigate the emergence of multiple transport timescales in the dissipative dynamics of a spin impurity coupled to a small, spatially disordered network of spins. 
Using a two-dimensional tight-binding model with dipolar interactions and local dephasing, we demonstrate that geometric heterogeneity leads to  hierarchical coupling strengths. This results in pronounced separation of dynamical timescales, far beyond those expected in homogeneous environments. By analyzing different metrics for dynamics, we identify distinct relaxation timescales associated with cluster-level equilibration and global equilibration. A minimal three-site model reveals the physical origin of the longest timescale: strong internal hybridization internally generates an effective detuning that suppresses transfer to other weakly-coupled sites, yielding a parametrically-enhanced relaxation time in the weak-dephasing regime. We corroborate this picture with nonequilibrium steady-state transport calculations and extensive simulations of disordered spin configurations, demonstrating orders-of-magnitude slowing of relaxation when hierarchical couplings are present. Our results highlight the central role of geometry and connectivity in spin networks and open quantum systems in general, and provide experimentally relevant predictions for relaxation times in small spin baths.
\end{abstract}

\maketitle

\section{Introduction}

Efficient transport phenomena, including the transfer of charge carriers, excitations, spins and quantum states, play a central role in a broad range of systems and applications, from energy harvesting \cite{Creatore2013,Fruchtman2016,DeSio2017,Rouse2019,Cavassilas2020,Hu2021QDTransport} to quantum information processing \cite{state0,state1,state2}. 
Such transport in disordered nanoscale systems can be optimized through the interplay of coherent and incoherent processes. Operating in this regime mitigates localization effects and can enable transport that surpasses classical diffusive limits. Consequently, a substantial body of theoretical work has focused on using incoherent effects to enhance quantum transport in quasi one-dimensional systems
\cite{Alan08,Plenio08,Plenio09,Plenio10,Plenio12,Cao13,Plenio21,mohseni2014energy,trautmann2018}
\cite{ZH2018,ZH2020,cygorek2022,kurt2023,Cao09,Maggie25,Rebentrost2009,Coates2021,Coates2023,Lawrence2026arxiv}
In higher-dimensional settings, related phenomena have also been explored. For instance, spin diffusion of a nitrogen vacancy (NV) center strongly coupled to a dipolar spin ensemble has been studied experimentally (see, e.g., Ref.~\cite{NVdiffusion21}). Exciton transport in a small network with a spherical geometry was shown to exhibit nontrivial transfer dynamics, with optimal performance occurring at distinct regimes of the decoherence rate \cite{Eriktwin23}. In the context of superconducting qubits, the relaxation time of a qubit coupled to randomly distributed resonant defects has been analyzed in Ref.~\cite{Dipti22}.

In heterogeneous systems, the thermalization dynamics often unfolds on multiple timescales. These timescales can be identified from the system's dynamical evolution and, theoretically, are encoded in the spectral structure of the Liouvillian, the generator of open-system dynamics \cite{breuer2007_decoherence,nitzan2013chemical,Manzano_2020,Rivas_2011}. The real parts of the Liouvillian eigenvalues determine the dissipative timescales: fast and slow relaxation processes correspond to eigenvalues with large and small negative real parts, respectively. The gap between the zero eigenvalue, whose corresponding eigenmode is the steady state of the dissipative dynamics, and the eigenvalue with the next smallest real part is known as the Liouvillian gap and sets the asymptotic relaxation timescale \cite{Cai_2013,Bonnes_2014,Marko_2015,Mori_2020,Mori_2023,Xiao_2026,Zhou_2022,Zhou_2026,Marko_2015}. Its closure signals the onset of dissipative phase transitions \cite{Kessler_2012,Fitzpatrick_2017,DPT18,Haga_2024,Mori_2024,Minganti_2018,Fink_2018,Young_2020,Shibata_2019}.

A separation of timescales typically reflects an underlying separation of energy scales in the system. Recent studies of local random Liouvillians have shown that separated clusters in the Liouvillian spectrum can encode a hierarchy of relaxation timescales associated with different classes of relaxation modes or the number of quasi-particles involved in the incoherent evolution~\cite{WangPRL20,Hartmann_2024,Sommer_2021,Haga_Ueda_2023}. Furthermore, this separation of relaxation timescales allows us to engineer distinct relaxation pathways, which could lead to anomalous relaxation dynamics 
such as the quantum Mpemba effect \cite{Ares_2025,Joshi_2024,Zhang_2025,Liu_2024,Medina_2025,Chatterjee_2024,Chatterjee_2023,Wang_2024,Longhi_2025,Calabrese_2026,Ma_2025,Yu_2025,Li_2025}. In the present work, we encounter
timescale separation in a different setting: a deterministic local-dephasing
Liouvillian whose slow modes arise from geometric hierarchy in the
Hamiltonian couplings. More generally, separation of relaxation timescales can arise, for example, from quasi-degenerate energy levels in the model \cite{Brumer14,Brumer15,Merkli_2015,Dodin_2016,Gerry24,Ivander23}, or from strong system–bath coupling that generates increasingly dark states in the spectrum \cite{Brett25}. A similar segment structure in Liouvillian also arises in many-body systems at strong coupling \cite{Renyi21}. Related behavior was reported in Ref.~\cite{ErikPRX22}, where distinct choices of intra- and inter-site tunneling amplitudes give rise to multiple relaxation timescales.
%

Quantum dynamics in disordered networks concerns fundamental open questions, especially when coherent evolution competes with environmental noise. Key questions concern the following directions: 
(i) breakdown of localization in two dimension (2D) and the interplay of coherence effects, disorder and environmental noise \cite{anderson1958absence,evers2008_anderson_transitions,Schwartz2007anderson,Harris2017NatPhoton,Choi2016Science,AA2008Nature,Roati2008Nature}; (ii) the role of geometry and many-body interactions in relaxation dynamics \cite{Jonas14,Luitz17,Bordia17,Abanin19}; (iii) and rare vs typical, individual vs statistical behavior \cite{Huse14,Sarang16}. 

\begin{figure}[t!]
\includegraphics[width=0.9\linewidth]{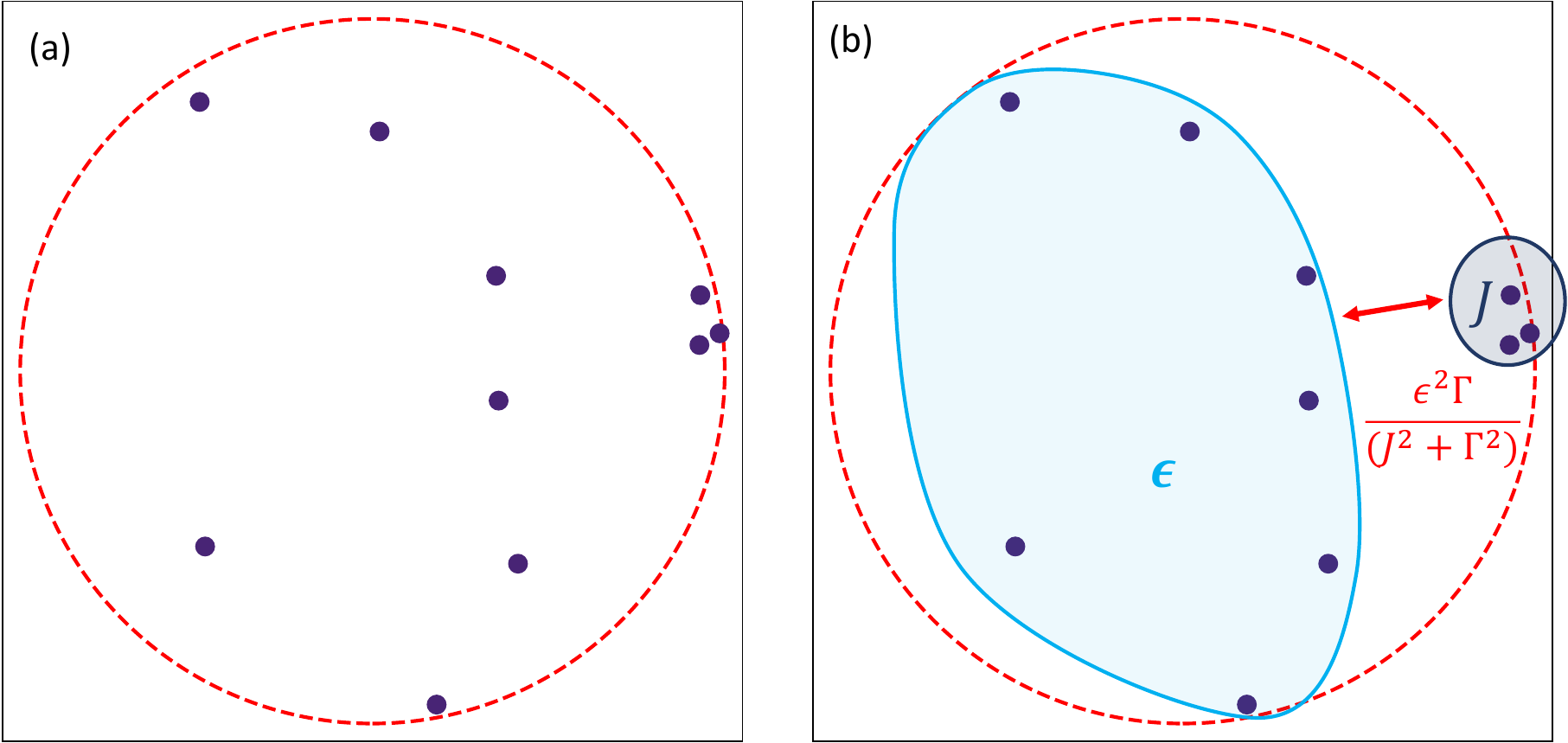}
\caption{
(a) Representative realization of a ten-site spin ensemble in two dimensions. 
All sites are subject to local dephasing. 
(b) Partitioning of the ensemble into two groups of a strongly-coupled (gray region) cluster and a weakly-coupled (light blue) group. 
These subnetworks (also referred to as clusters or small baths) are connected on the timescale calculated with a minimal model, Sec. \ref{sec: Minimal model analysis: Timescale Hierarchies}, reflecting the underlying geometric hierarchy of the system.
}
\label{fig:scheme}
\end{figure}

In this work, we address the fundamental problem of quantum equilibration dynamics in heterogeneous two-dimensional quantum systems where coherent dynamics coexist with incoherent local dephasing. Motivated by experiments with NV centers \cite{Awsch13}, we study a finite network of coupled sites [Fig. \ref{fig:scheme}(a)] that provides a minimal yet versatile setting for exploring the competition between coherence, disorder, and decoherence in quantum transport.
Such a network could represent a single NV center coupled to a collection of nearby spins. As such, in Fig. \ref{fig:scheme}(a) one site would be designated as the NV center, while the remaining sites constitute a small spin bath. 
To account for additional environmental effects, e.g., lattice phonons, many-body spin interactions, and other sources of noise, we introduce local dephasing on each spin. 
We focus on the single-excitation regime. As such, we analyze relaxation dynamics ($T_1$) as well as the properties of the resulting steady state.


We address the following open questions: 

(i) {\bf Emergence of timescales:} 
What mechanisms give rise to single-step versus multistep relaxation dynamics in disordered networks? How do geometric heterogeneity and hierarchical coupling strengths generate widely-separated dynamical timescales, and how are these reflected in dynamical and Liouvillian spectral diagnostics?

(ii) {\bf Transport pathways and their competition:}
What determines the dominant relaxation pathways in small disordered networks? How does environmental dephasing control pathway selection and switching between them?

(iii) {\bf Geometry-dynamics relations:}
How does the spatial structure of the network, such as clustering and connectivity, translate into effective energy scales, bottlenecks, and coupling-generated detuning effects that govern transport and relaxation?

(iv) {\bf Optimized transport:} 
Can distinct dephasing regimes lead to different optimal transport conditions? Can heterogeneous networks exhibit multiple transport optima associated with competing pathways?

(v) {\bf Ensemble averages:}
To what extent are the observed dynamical features, such as hierarchical timescales and pathway competition, robust across different disorder realizations, and what statistical signatures characterize these effects?

By analyzing two representative configurations in detail, we address questions (i)–(iv) and uncover two defining characteristics of heterogeneous 2D systems: pronounced hierarchies of relaxation timescales and the competition between multiple distinct relaxation pathways.
For a fixed spin density, we further analyze question (v), the scaling of ensemble-averaged properties with the geometric size of the system and demonstrate robust scaling behavior. 

The paper is organized as follows. In Sec.~\ref{sec: Physical Setup, Model and dynamical Measures}, we introduce the physical setup, the equations of motion, and the dynamical measures used throughout the study. To illustrate the different timescales in the problem, Sec.~\ref{sec: Minimal model analysis: Timescale Hierarchies} analyzes a minimal three-site model, identifying relaxation timescales at both high and low dephasing rates. We then examine two representative configurations in detail. Configuration A, discussed in Sec.~\ref{sec:ConfA}, allows us to examine the hierarchical nature of the relaxation dynamics. Configuration B, analyzed in Sec.~\ref{sec:ConfB}, illustrates the presence of competing relaxation pathways and the role of dephasing in determining which pathway dominates. In Sec.~\ref{sec:hierarchical_statistics} we introduce statistical signatures of the previously studied effect in hierarchical networks, and in Sec.~\ref{sec:ensemble}, we extend our analysis to an ensemble of configurations, exploring the generality of the observed behavior. Finally, Sec.~\ref{sec:summary} presents our conclusions.

\section{Physical Setup, Model and dynamical Measures}
\label{sec: Physical Setup, Model and dynamical Measures}

\subsection{Physical setup}


This study addresses general questions associated with transport in disordered systems, with relevance to a broad range of physical platforms and realistic scenarios. Nevertheless, it is instructive to consider a particularly relevant experimental realization consisting of a single NV center coupled to a small spin environment. Such systems are implemented in diamond samples implanted with a low density of nitrogen ions, enabling individual NV centers to be optically addressed using a confocal microscope. This experimental platform provides a versatile testbed for addressing different aspects in the field of open many-body quantum systems: single spin open quantum system dynamics and control \cite{deLange2010,Ryan2010,Naydenov2011}; single spin cooling or sensing applications \cite{Balas8}; two-NV spin dynamics and control \cite{Neumann08,Bernien13}; bath-mediated synchronization \cite{Brenes24}; quantum correlations and quantum information transfer \cite{BarGill2012, BarGill2013,Pham2012,Romach2015,Farfurnik2015}. 

We refer to the implanted nitrogen ions as a ``small bath" or ``spin network". In terms of energy parameters, we assume long range dipolar couplings between all nitrogen impurities, including both the NV center and the surrounding bath spins. We further consider the NV center to be energy-resonant with the bath \cite{NirDress13}. 
We include a local dephasing effect for each spin impurity, representing the impact of e.g., phonons, or additional spin baths which are not explicitly incorporated into the Hamiltonian we build for our spin system. Referring to Fig.~\ref{fig:scheme}, one of the blue dots represents the NV center, while the others correspond to the surrounding spin network. However, since all spins are taken to be resonant, any given impurity can equivalently play the role of the NV, with the remaining spins forming the environment.
Consequently, Fig.~\ref{fig:scheme} can represent ten distinct realizations of the setup, in each of which a different impurity is selected to serve as the NV, which is initially prepared in the excited state. 

The dipolar coupling between two electronic spins separated by a distance $r$ is given by
\begin{equation}
J = \frac{\mu_0}{4\pi} \frac{(g_e \mu_B)^2}{h\, r^3}
\left(1 - 3\cos^2\theta\right),
\label{eq:Jspin}
\end{equation}
where $\theta$ is the angle between the inter-spin vector and the external magnetic field (or the intrinsic quantization axes of the spins).
Evaluating this expression for the separation of $r \simeq 0.5\,\mathrm{nm}$ and $\theta = 0^\circ$ yields a coupling strength of order
$J \sim 1\,\mathrm{GHz}$.
In simulations, we therefore normalize one unit of distance to correspond to $0.5\,\mathrm{nm}$.
With this choice, setting $J_{\rm max}=1$ fixes the time unit via
%
$t_0 = \frac{1}{J_{\max}}$.
%
One simulation time unit thus corresponds to approximately $t_0 \simeq 1\,\mathrm{ns}$ in physical units.
In particular, we consider a regime corresponding to approximately $100$ spins within a radius of
$50\,\mathrm{nm}$. Interpreting this number as an effective three dimensional concentration,
and using the atomic density of diamond,
$n_{\mathrm C}\simeq 1.76\times10^{23}\,\mathrm{cm^{-3}}$,
this corresponds to a bulk spin concentration of order
$\sim 10\,\mathrm{ppm}$.
This concentration is consistent with implantation doses of order $1\times10^{12}\,\mathrm{ions/cm^2}$ and lies within the experimentally relevant range for realistic implanted nitrogen spin baths in diamond.

The analysis in this work corresponds to a two dimensional spin network. This choice is motivated by the fact that, in many realizations of implanted spin systems, the relevant spins occupy a shallow layer near the diamond surface, whose thickness is much smaller than its lateral extent. Ion implantation energies of up to a few tens of keV typically produce defect layers confined to depths of only a few to a few tens of nanometers \cite{PhysRevB.93.035202}. Such ensembles therefore form an effectively planar distribution of spins, where the vertical spread is small compared with the typical lateral separation between spins. In this regime, the dominant spatial structure of the dipolar couplings is well captured by a two dimensional geometry. Modeling the system in two dimensions thus provides a realistic description of the effective spin network while simplifying the analysis.

\subsection{Hamiltonian and equation of motion}

We consider a spin network with $N$ sites and a single excitation in the system, i.e., all spins are in the ground-state except for one that is excited (``the NV"). The Hamiltonian of the ensemble of spins is given by a tight-binding model,
\begin{equation} \label{eq:ham}
    \hat H = \sum_{i=1}^{N} E_i \ket{i} \bra{i} + \sum_{i \neq j} J_{ij} \ket{i} \bra{j}.
\end{equation}
Here, $E_i$ is the energy of site \textit{i}, and \(J_{ij}\) is the tunneling energy between sites \textit{i} and \textit{j} (dipolar coupling energy). In the context of excitation transfer, the diagonal elements represent the single-excitation energies of each spin, while $J_{ij}$ denotes the flip-flop (exchange) coupling between spins $i$ and $j$. 
For simplicity, the off-diagonal elements are chosen to be real and positive.  We use a dipolar power law function to describe the tunneling energies,
\begin{equation} \label{powerlaw}
     J_{ij} = \frac{J_{\rm max}}{|r_i - r_j|^3},
\end{equation}
with $J_{\rm max}$ discussed below Eq. (\ref{eq:Jspin}).
One of the sites in the ensemble will be considered as the NV, while the rest will serve as the small spin bath. In simulations, we assume resonant conditions, thus we set all energies to $E_i=0$.

In Fig.~\ref{fig:scheme}(a) we demonstrate a randomly generated configuration consisting of ten sites. We begin by examining the geometry of this setup.  
Notably, three sites are located in close spatial proximity; this cluster is highlighted by a gray circle in Fig.~\ref{fig:scheme}(b). 
These three sites are, in turn, coupled to a surrounding layer of approximately two to three additional sites, which themselves couple to the remaining sites in the system. We group these surrounding sites within the light-blue domain.

\begin{figure}[t!]
\includegraphics[width=0.9\linewidth]{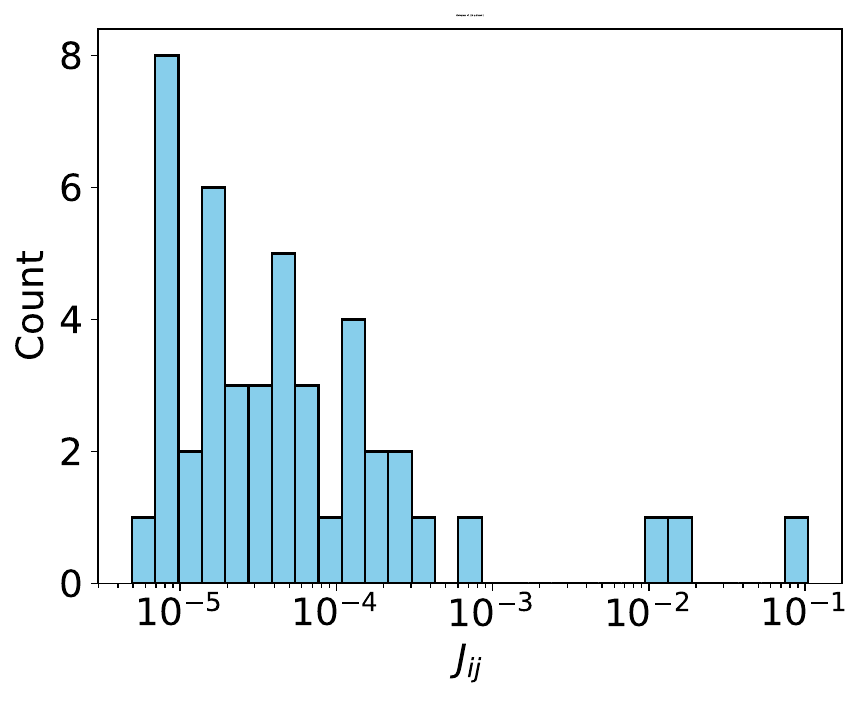}
\caption{Histogram of tunneling couplings for the geometry presented in Fig.~\ref{fig:scheme}. 
Two well-separated coupling scales are identified: values on the order of $J_{ij}\sim 10^{-2}-10^{-1}$, and smaller values spanning $10^{-5}-10^{-3}$. The former are associated with the three sites (gray ensemble) in close proximity to one another.
Smaller $J_{ij}$ values correspond to all other tunneling elements between sites separated by distances comparable to the system's radius.
}
\label{fig:Jij}
\end{figure}

We refer to each of the gray and light-blue domains as a \emph{mini-bath} or \emph{subnetwork}. This partitioning is supported by a quantitative analysis of the tunneling amplitudes: In Fig.~\ref{fig:Jij} we present a histogram of the coupling strengths $J_{ij}$. 
The three largest values of $J_{ij}$ correspond to the intra-cluster couplings within the three-site group of the gray domain. 
The remainder of the distribution comprises couplings between the two spin subclusters, as well as couplings internal to the blue spin network.
These two clusters of coupling energies are separated by about two orders of magnitude.

We assume that the dephasing environment acts locally and in an uncorrelated manner on each spin. The local Lindblad Quantum Master Equation (QME) for the $N$-site density matrix $\hat \rho$ is given by ($\hbar\equiv 1$) \cite{breuer2007_decoherence},
\begin{equation} 
\label{lindpd}
    \dot{\hat \rho} = -i\left[\hat H, \hat \rho\right] + \sum_{j=1}^N \Gamma_j \left(\hat L_j \hat \rho \hat L_j^\dagger - \frac{1}{2} \left\{\hat L_j^\dagger \hat L_j, \hat \rho\right\} \right).
\end{equation}
In a compact form,
$\dot{\hat \rho} = \hat{ \mathcal L} \hat\rho(t)$, where the Liouvillian $\hat{\mathcal L}$ includes both unitary and nonunitary terms.

Note that the density matrix follows the evolution of all spins: NV and all other spins, while additional local dephasing effects are captured by the dissipator.
The Lindblad jump operators are given by $ \hat L_j = \ket{j} \bra{j}$, with $\Gamma_j$ as dephasing rate constants. The dynamics evolves in the site-local basis. For simplicity, we assume that all $\Gamma_j$ are identical. The steady state solution of this equation is a state with zero coherences and equal population on every site, $\rho_{jj}(t\to \infty)=1/N$, with $N$ the total number of sites in the system. Thus, for a fully connected 10-site system, at equilibrium each site's population is at 10\%.
The central questions that we address in this study concern the relaxation times and pathways leading to this equilibrium state. In the next subsection, different measures used to quantify the dynamics and steady state of the system are introduced.

\subsection{Measures for relaxation and transfer processes}
\label{subsec: Measures for relaxation and transfer processes}

We investigate the dynamics of a single excitation initially localized on a specific site $i$, with $\hat \rho(0)=\ket{i}\!\bra{i}$. The resulting nonequilibrium evolution and its pathways is characterized through several complementary observables. We further consider a nonequilibrium steady-state configuration which, as demonstrated in Sec.~\ref{sec:time2}, reveals additional information about the competing transport pathways.

{\bf i. Population dynamics.}
The direct measure for relaxation timescales is the evolution of populations at each site, $P_j(t)=\rho_{jj}(t)$, $j=1,2,...,N$ with $N$ sites, from a given initial condition. 

{\bf ii.\ Purity.} 
The purity of a quantum state is defined as $D(t) =\mathrm{Tr}\!\left[\hat\rho(t)^2\right] $, and is bounded from above and below $\frac{1}{N}\leq D(t) \leq 1$ \cite{NielsenChuang2000}. The upper bound corresponds to a pure state, whereas the lower bound corresponds to a completely mixed state. The relaxation dynamics therefore drive the system from an initially pure state toward a completely mixed state. 

{\bf iii. Liouvillian spectrum.}
The eigenvalues of the Liouvillian encode information about the relaxation timescales: the inverse of the real 
part of a nonzero eigenvalue sets the corresponding decay time, while the zero eigenvalue corresponds to the steady state.
Furthermore, the projections of the Liouvillian eigenvectors onto the site basis reveal how the different relaxation modes are distributed spatially across the system.
This analysis is delegated to Appendix \ref{AppA}.

{\bf iv.  Integrated survival and transfer time measures.}
We suggest time-integrated population measures, related to the residence time and mean first passage time (see Appendix \ref{AppA}).
Starting from an initial condition $\rho(0)=|i\rangle\langle i|$, we define the measure
\begin{equation}
T_{ij}=\int_0^\infty \rho_{jj}(t)\,dt,
\label{eq:Tij}
\end{equation}
excluding the steady state, as we explain next.
Here, $\rho_{jj}(t)$ is the population at site $j$ at time $t$.
In a vectorized form, $\rho(t)=e^{\hat {\mathcal L} t}\rho(0)$, and the time integrated population can be written as
\begin{equation}
T_{ij}=[\hat {\mathcal  L}^{-1}\rho_{ii}]_{jj},
\end{equation}
Here, $\hat {\mathcal L}^{-1}$ denotes the pseudo inverse of the Liouvillian, defined by {\it excluding} the steady state mode, see Appendix \ref{AppA}.
This construction regularizes the time integral, which would diverge otherwise. 

The diagonal elements, $T_{ii}$, quantify the {\it survival} or residence time of an excitation prepared at site $i$, to be found at the same site during the relaxation process. 
Off-diagonal elements $T_{ij}$ $(i\neq j)$ quantify the transfer time of excitations from site $i$ (initial condition) to site $j$ during the relaxation process.
Together, the matrix $T_{ij}$ provides a compact description of how relaxation unfolds spatially across the network. 

While $T_{ij}$ are not mean first-passage times (MFPT) in the strict probabilistic sense, they provide a robust diagnostic of population relaxation and transfer in heterogeneous systems, particularly when multiple timescales are present.  
Since the steady-state contribution is removed, $T_{ij}$ can take negative values and should be interpreted as a relative, time-integrated population measure rather than a strictly positive timescale.  
More specifically, because the steady-state baseline ($1/N$) is subtracted from the definition, sites that remain \emph{underpopulated} throughout the relaxation process, compared to the steady state value, contribute a net {\it negative} integrated area.
Formal relations to trapping-based MFPT definitions and spectral representations are discussed in Appendix \ref{AppA}.

{\bf v. Nonequilibrium steady state: population flux.}
The dynamical analysis is complemented by reformulating the problem as a nonequilibrium steady-state setup. This is achieved by introducing a loss process at a designated ``final" site \( f \), described by the jump operator
\begin{equation}
\hat L_l = \sqrt{\gamma_l}\, |0\rangle\langle f| ,
\end{equation}
with a rate constant \( \gamma_l \), while enforcing conservation of excitations within the system. 
In this framework, excitations enter the system at the entrance site $0$ with a constant flux, and exit at site $f$ with the same rate, establishing a nonequilibrium steady transport process. 
The resulting population flux is defined as
\begin{equation}
\eta = \gamma_l \rho^{\mathrm{NESS}}_{ff},
\end{equation}
where $\rho^{\mathrm{NESS}}_{ff}$ denotes the steady-state population of site $f$ (see Ref.~\cite{Maggie25,Lawrence2026arxiv} for recent implementations). Directly associating the inverse of the flux with a relaxation timescale is possible in some cases, but is generally nontrivial \cite{Kalantar19}. In practice, the loss rate constant $\gamma_l$  must be chosen sufficiently small to not dominate or mask intrinsic decoherence processes, but not too small that it becomes the rate-determining step. 
This intermediate regime allows the ``natural'' dynamics of the model to be observed, as we demonstrate in Sec.
\ref{sec: Minimal model analysis: Timescale Hierarchies}.

\section{Minimal model analysis: Timescale Hierarchies}
\label{sec: Minimal model analysis: Timescale Hierarchies}

\begin{figure}[t!]
\includegraphics[width=0.9\linewidth]{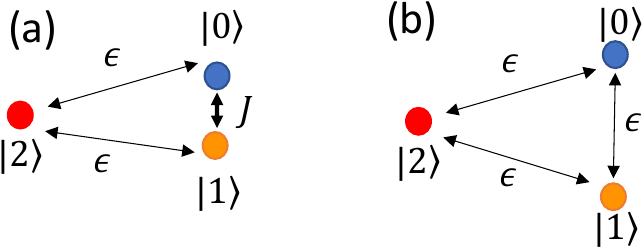}
\caption{Scheme of a minimal three-level model for extraction of relaxation lifetime. Excitation is prepared at site $|0\rangle$ and the timescale for it to transfer to site $|2\rangle$ is analyzed. All sites experience local dephasing at strength $\Gamma$. 
(a) Hybridized pair and remote site: Sites $|0\rangle$ and $|1\rangle$ are strongly connected by a tunneling energy $J$, and both connected to site $|2\rangle$ with a weaker coupling $\epsilon$.
(b) Homogeneous system:
All sites are connected by tunneling couplings of the same order of magnitude, $\epsilon$.
}
\label{fig:H3}
\end{figure}

We begin by analyzing a minimal three-site model in order to identify the physical origins of the distinct timescales governing the dynamics in our system. 
The central result of this Section is a minimal physical picture explaining the emergence of the long survival timescale, Eq.~(\ref{eq:ttr}) below, observed in our simulations. This timescale arises from the interplay between strong internal hybridization of spins ($J$), weak coupling to a spatially remote site ($\epsilon$), and local dephasing ($\Gamma$).

We consider a three-site model described by the Hamiltonian
\begin{equation}
\hat H =
\begin{pmatrix}
0 & J & \epsilon \\
J & 0 & \epsilon \\
\epsilon & \epsilon & 0
\end{pmatrix},
\label{eq:H3}
\end{equation}
where sites $|0\rangle$ and $|1\rangle$ are assumed spatially close, thus strongly coupled with tunneling strength $J$, while site $|2\rangle$ is more distant and weakly coupled to the others, with tunneling strength $\epsilon$, such that $J \gg \epsilon$.

All sites experience local dephasing at a rate $\Gamma$, representing coupling to a phonon bath or to $^{13}\mathrm{C}$ nuclear spins.  
Importantly, no external detuning is applied; as we show next, an energy mismatch arises internally from the hybridization scale $J$. A schematic representation of the three-site configuration is shown in Fig.~\ref{fig:H3}, where the heterogeneous coupling setup contrasts with its homogeneous counterpart.

\subsection{Derivation of the effective relaxation timescale}
\label{sec:time1}

In the limit $J \gg \epsilon$, it is convenient to work in the symmetric and antisymmetric basis of the strongly coupled pair,
\begin{equation}
|+\rangle = \frac{1}{\sqrt{2}} (|0\rangle + |1\rangle), 
\qquad
|-\rangle = \frac{1}{\sqrt{2}} (|0\rangle - |1\rangle).
\end{equation}
In the basis $\{|+\rangle, |-\rangle, |2\rangle\}$, the Hamiltonian becomes
\begin{equation}
\hat H' =
\begin{pmatrix}
J & 0 & \sqrt{2}\,\epsilon \\
0 & -J & 0 \\
\sqrt{2}\,\epsilon & 0 & 0
\end{pmatrix}.
\label{eq:Hp}
\end{equation}
The antisymmetric state $|-\rangle$ (energy -$J$) is completely decoupled from the site $|2\rangle$ and therefore forms a dark state. Only the symmetric state $|+\rangle$ (energy $J$) is coupled to the remote site, with matrix element $V = \langle 2 | \hat H | + \rangle = \sqrt{2}\,\epsilon$, and an energy difference set by $J$.

In the Markovian limit, the effect of dephasing is incorporated into the transition probability as a  broadening of the energy. This is described by a Lorentzian density of states,
$D(E) = \frac{1}{\pi} \frac{\Gamma}{E ^2 + \Gamma^2}$,
with $E$ being the Hamiltonian energy difference between the initial and final states.  
We apply Fermi’s Golden Rule to the off-resonant transition from $|+\rangle$ to $|2\rangle$, $\gamma_{\mathrm{tr}} = 2\pi |V|^2 D(J)$ and identify the transition element $|V|^2=2\epsilon^2$ from Eq. (\ref{eq:Hp}). This yields an effective transfer rate ($\gamma_\mathrm{tr}$),
%
\begin{equation}
\gamma_{\mathrm{tr}} 
= \frac{4 \epsilon^2 \Gamma}{J^2 + \Gamma^2},
\end{equation} 
where the corresponding transfer timescale to the target site is therefore
\begin{equation}
t_{\mathrm{tr}} \sim \gamma_{\mathrm{tr}}^{-1}
= \frac{J^2 + \Gamma^2}{4 \epsilon^2 \Gamma}.
\label{eq:ttr}
\end{equation}
In the weak dephasing regime, $\Gamma \ll J$, this reduces to
\begin{equation}
t_{\mathrm{tr}} \sim \frac{J^2}{\epsilon^2 \Gamma},
\label{eq:newtime}
\end{equation}
while in the strong dephasing regime, $\Gamma \gg J$ one finds
\begin{equation}
t_{\mathrm{tr}} \sim \frac{\Gamma}{\epsilon^2}.
\end{equation}

\begin{table*}[t]
\caption{
Representative numerical relaxation timescales corresponding to the scalings derived in Sec.~\ref{sec:time1}. We compare hierarchical connectivity, consisting of a strongly hybridized pair weakly coupled to a remote site,
with homogeneous ensemble (bath) couplings in the weak and strong dephasing regimes. 
}
\label{tab:numerical_timescales}
\centering
\renewcommand{\arraystretch}{2.1}

\begin{tabular}{lcc}
\hline\hline
System / connectivity & Relaxation time & Numerical value \\ 
\hline
\multicolumn{3}{c}{\textit{(a) Weak dephasing regime $\Gamma\ll J$: } 
$J = 10^{-2},\; \epsilon = 10^{-4},\; \Gamma = 10^{-5}$} \\
\hline
Hybridized pair + remote site 
& $t_{\mathrm{tr}} \sim \dfrac{J^{2}}{\epsilon^{2}\Gamma}$ 
& $\sim 10^{9}$ \\
Homogeneous baths (coupling $J$) 
& $t_{\mathrm{tr}} \sim \dfrac{1}{\Gamma}$ 
& $\sim 10^{5}$ \\
Homogeneous bath (coupling $\epsilon$) 
& $t_{\mathrm{tr}} \sim \dfrac{1}{\Gamma}$ 
& $\sim 10^{5}$ \\
\hline
\multicolumn{3}{c}{\textit{(b) Strong dephasing regime $\Gamma\gg J$: } 
$J = 10^{-2},\; \epsilon = 10^{-4},\; \Gamma = 10^{-1}$} \\
\hline
Hybridized pair + remote site 
& $t_{\mathrm{tr}} \sim \dfrac{\Gamma}{\epsilon^{2}}$ 
& $\sim 10^{7}$ \\
Homogeneous bath (coupling $J$) 
& $t_{\mathrm{tr}} \sim \dfrac{\Gamma}{J^{2}}$ 
& $\sim 10^{3}$ \\
Homogeneous bath (coupling $\epsilon$) 
& $t_{\mathrm{tr}} \sim \dfrac{\Gamma}{\epsilon^{2}}$ 
& $\sim 10^{7}$ \\
\hline\hline
\end{tabular}
\end{table*}

The identification of a {\it long} transfer timescale Eq. (\ref{eq:newtime}) between strongly-coupled sites ($J$ tunneling) to weakly-coupled sites ($\epsilon$ tunneling) is an important observation in our context, which we reproduce in simulations and show to lead to hierarchal relaxation dynamics and competing relaxation pathways. This effect can be rephrased as follows: when two sites are in close proximity (``twins''), they strongly hybridize with each other, effectively ``refusing'' to accept or transfer excitations to sites outside their strongly coupled group.  
We refer to this as the {\it ``twin fortress''} effect, when two sites are involved, but a strongly-tight subcluster could involve more than two sites and show the ``fortressing" effect. 
It can be overcome when local dephasing is sufficiently strong, on the order of the energetic barrier $J$, facilitating excitation transfer between the strongly coupled group and the rest of the network.

It is also useful to compare the heterogeneous model with homogeneous scenarios [Fig. \ref{fig:H3}(a) and (b)]. In this case, all sites are coupled via comparable tunneling elements, which can be large ($J$) or small ($\epsilon$). Table~\ref{tab:numerical_timescales} 
summarizes the transfer timescales in the limits of weak and strong dephasing for different connectivity structures.
For homogeneous coupling cases, the excitation is always resonantly coupled between sites. As a result, the weak dephasing regime is governed by the timescale $1/\Gamma$, while strong dephasing leads to a Zeno-like slowdown controlled by the square of the coupling strength.
In contrast, in the configuration of a hybridized pair plus remote site, strong internal coupling internally generates an effective detuning. 
In the weak dephasing regime, this produces the significantly longer timescale $J^2/(\epsilon^2 \Gamma)$. Only when the dephasing becomes sufficiently strong does the system cross over to incoherent transfer governed by $\epsilon$.

This comparison demonstrates that the emergent timescale identified in this work is not a trivial extension of homogeneous ensemble physics. Instead, it arises as a consequence of strong internal hybridization combined with weak external connectivity and dephasing effects. 

To illustrate the emerging long timescale, consider the weak dephasing limit with $\Gamma = 10^{-6}$. We further assume that $J\sim 0.1$ while $\epsilon=10^{-4}$. In the homogeneous case, with tunneling energies $J$ or $\epsilon$, the relaxation dynamics will be dictated by the slow process, with timescale $1/\Gamma = 10^6$. By contrast, in the hierarchical-coupling case, when the system is initialized within one of the strongly coupled pairs, the relaxation dynamics is further suppressed by a factor $(J/\epsilon)^2=10^6$. Thus, the {\it hierarchical} structure of the tunneling amplitudes leads to {\it orders-of-magnitude slowing down of the relaxation dynamics}, as we show in Table \ref{tab:numerical_timescales}.



\subsection{Nonequilibrium steady state analysis}
\label{sec:time2}

The emergence of a long lifetime in heterogeneous systems, which we identified in Sec. \ref{sec:time1}, can also be derived in the steady state limit, by formulating the problem in a nonequilibrium steady state setting. 

We again consider the Hamiltonian in Eq.~(\ref{eq:H3}), which consists of three sites with hierarchical couplings $J \gg \epsilon$.
The system is now driven out of equilibrium by introducing a decay process (loss) at site $2$ with a rate constant $ \gamma_l$, while enforcing the conservation of a single excitation: a constant flux of particles is injected at entrance site $0$, and extracted at site $2$ at the same rate, establishing a nonequilibrium steady-state. The population flux is defined as
$\eta = \gamma_l \rho^{\mathrm{NESS}}_{22}$, where \( \rho^{\mathrm{NESS}}_{22} \) denotes the steady-state population of site $2$.

Following Refs.~\cite{Kalantar19,Maggie25}, we associate the inverse of the flux with the relaxation timescale.  An exact solution of the steady state problem leads to 
\bea
&&\eta =
\nonumber\\
&&\frac{4\Gamma\gamma_l(2\Gamma+\gamma_l)\epsilon^2}
{4\Gamma^3\gamma_l + 4\gamma_l^2\epsilon^2 + 4\Gamma^2(\gamma_l^2+6 \epsilon^2) + \Gamma[\gamma_l^3+4\gamma_l(J^2 +5 \epsilon^2)]}.
\nonumber\\
\label{eq:eta}
\eea
Note that the system reaches a nonequilibrium steady state that depends on the value of $\gamma_l$, as well as on other parameters of the problem. 

We now assume the following hierarchy of energy scales, $ J \gg \epsilon \gg \Gamma \gg \gamma_l$. 
We further require
$ \frac{J}{\epsilon} \gg \sqrt{\frac{\Gamma}{\gamma_l}}$.
This second condition ensures that the internal system timescale,
$J^2/(\epsilon^2 \Gamma)$ is longer than the leakage timescale, $ 1/\gamma_l$.
Under these conditions, we obtain
\bea
\eta \approx  \frac{\Gamma\epsilon^2}{J^2}.
\eea
%

In the opposite strong-dephasing regime, $\Gamma \gg J \gg \epsilon \gg \gamma_l$, 
one retains in the denominator of Eq.~(\ref{eq:eta}) only the leading term 
$\Gamma^3 \gamma_l$, yielding
\begin{equation}
\eta \propto \frac{\epsilon^2}{\Gamma}.
\end{equation}
This result was derived while further requiring that 
$ \Gamma^3 \gamma_l \gg \Gamma^2 \epsilon^2$, 
[see denominator of Eq.~\eqref{eq:eta}],
or equivalently 
$\Gamma \gg \epsilon^2/\gamma_l$. 
This condition ensures that the leakage timescale $ 1/\gamma_l$ is shorter than the characteristic timescale of the internal dynamics, $\Gamma/\epsilon^2$. 

In both strong and weak dephasing limits, we therefore ensure that $\gamma_l$ is not chosen ``too'' small, in the sense that it becomes the rate-determining step controlling the observed trends;  when $\gamma_l$ is rate limiting, the population flux reduces to $\eta = \gamma_l/3 $,  corresponding to a steady state with an equal population distribution among the three sites.  We deliberately avoid this regime as it obscures the dependence on the physically relevant system parameters.

Organizing these results, we get from Eq.~\eqref{eq:eta} that the inverse of the flux, which can be regarded as a timescale for the transfer process, is
\bea
\eta^{-1}\approx
\begin{cases}
\frac{\Gamma}{\epsilon^2} &
{\rm for\,\,     } \Gamma\gg J\gg \epsilon, \,\,\,\gamma_l\gg \epsilon^2/\Gamma   \\
\frac{J^2}{\epsilon^2}\frac{1}{\Gamma} &
{\rm for\,\,     }  J\gg \epsilon\gg \Gamma, \,\,\, J^2\gamma_l\gg \epsilon^2\Gamma
\\
%
\end{cases}
\eea
in agreement with timescales identified in Table \ref{tab:numerical_timescales} for a hybridized pair + remote site, based on the Fermi Golden rule.

As the correspondence between dynamical measures and nonequilibrium steady state was demonstrated here,
in Secs. \ref{sec:ConfA}-\ref{sec:ConfB} we show that steady state calculations complement dynamical simulations, providing additional pathway insights.

\subsection{Discussion}

The minimal three-level model provides valuable insight into larger networks with $N$ sites, such as the $N=10$ system shown in Fig.~\ref{fig:scheme}. Inspection of the tunneling-coupling histogram in Fig.~\ref{fig:Jij} reveals a clear separation into two classes: strong couplings, with typical magnitude $J \approx 0.01$, and much weaker couplings characterized by an energy scale $\epsilon \approx 10^{-4}$, such that $J \gg \epsilon$. This  hierarchy of coupling strengths gives rise to an additional emergent energy (inverse-time) scale governing the effective coupling between the two subnetworks, originating from the interplay between coherent tunneling and local dephasing.

In the strong-dephasing regime  $\Gamma \gg J$,  dynamics initiated at the strongly-connected subnetwork (see Fig.~\ref{fig:scheme}(b))  will exhibit multiple timescales: a fast decay of coherences on a timescale $1/\Gamma$ (which for $\Gamma = 0.1$ is $t_{\rm{tr}} \approx 10$); local equilibration within the tightly coupled (gray) cluster on a timescale $ \Gamma/J^2 $, 
which for \( \Gamma = 0.1 \) and \( J \approx 0.01 \) yields \( t_{\rm{tr}} \sim 10^3 \); 
and global equilibration involving the remaining sites on a timescale $ \Gamma/\epsilon^2 $, 
which for \( \Gamma = 0.1 \) and \( \epsilon \approx 10^{-4} \) yields \( t_{\rm tr} \sim 10^7 \).

In contrast, in the weak dephasing regime, $ \Gamma \ll J $, 
a new, much longer relaxation timescale emerges,
$
t_{\rm tr} \sim \frac{J^2}{\epsilon^2}\frac{1}{\Gamma}$,
which reaches $ t_{\rm tr} \sim 10^{10}$ for 
\( J = 0.01 \), \( \epsilon = 10^{-4} \), and \( \Gamma = 10^{-6} \).
In Secs. \ref{sec:ConfA} and \ref{sec:ConfB} we illustrate these timescales on two representative configurations.


\section{Configuration A:  Hierarchal relaxation}
\label{sec:ConfA}

\begin{figure}[htbp]
\includegraphics[width=0.85\linewidth]{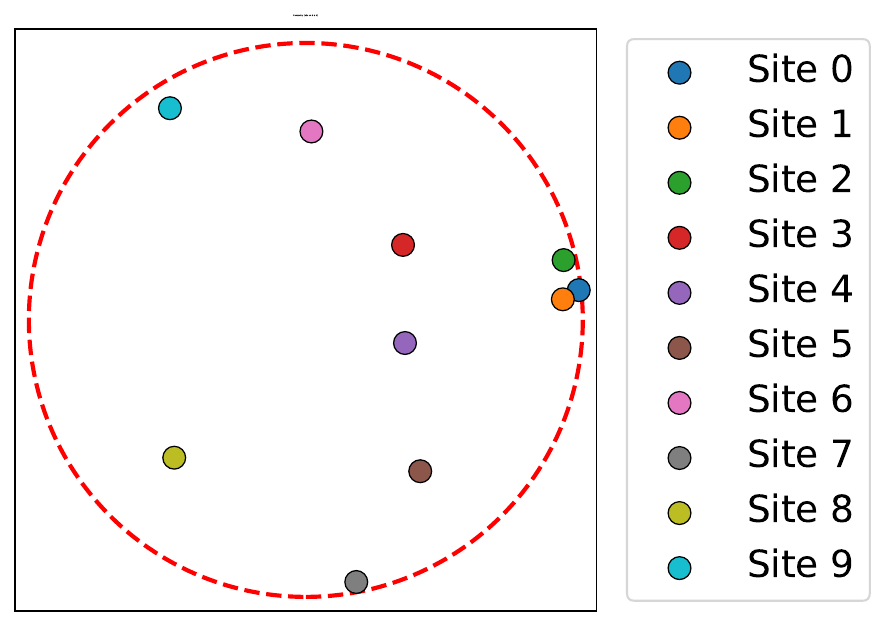}
\caption{
Configuration A.
Representative realization of a ten-site ensemble within a circular region. Sites are labeled according to the numbering used throughout the dynamical analysis.
}
\label{fig:geom1}
\end{figure}

We examine in detail a representative configuration that illustrates two key features: hierarchical relaxation and multiple competing relaxation pathways. Configuration A, shown in Fig.~\ref{fig:geom1}, is analyzed in terms of its relaxation dynamics, transfer and survival timescales, and steady-state behavior using several complementary measures.

\begin{figure*}[htpb]
\centering
\includegraphics[width=0.95\textwidth]{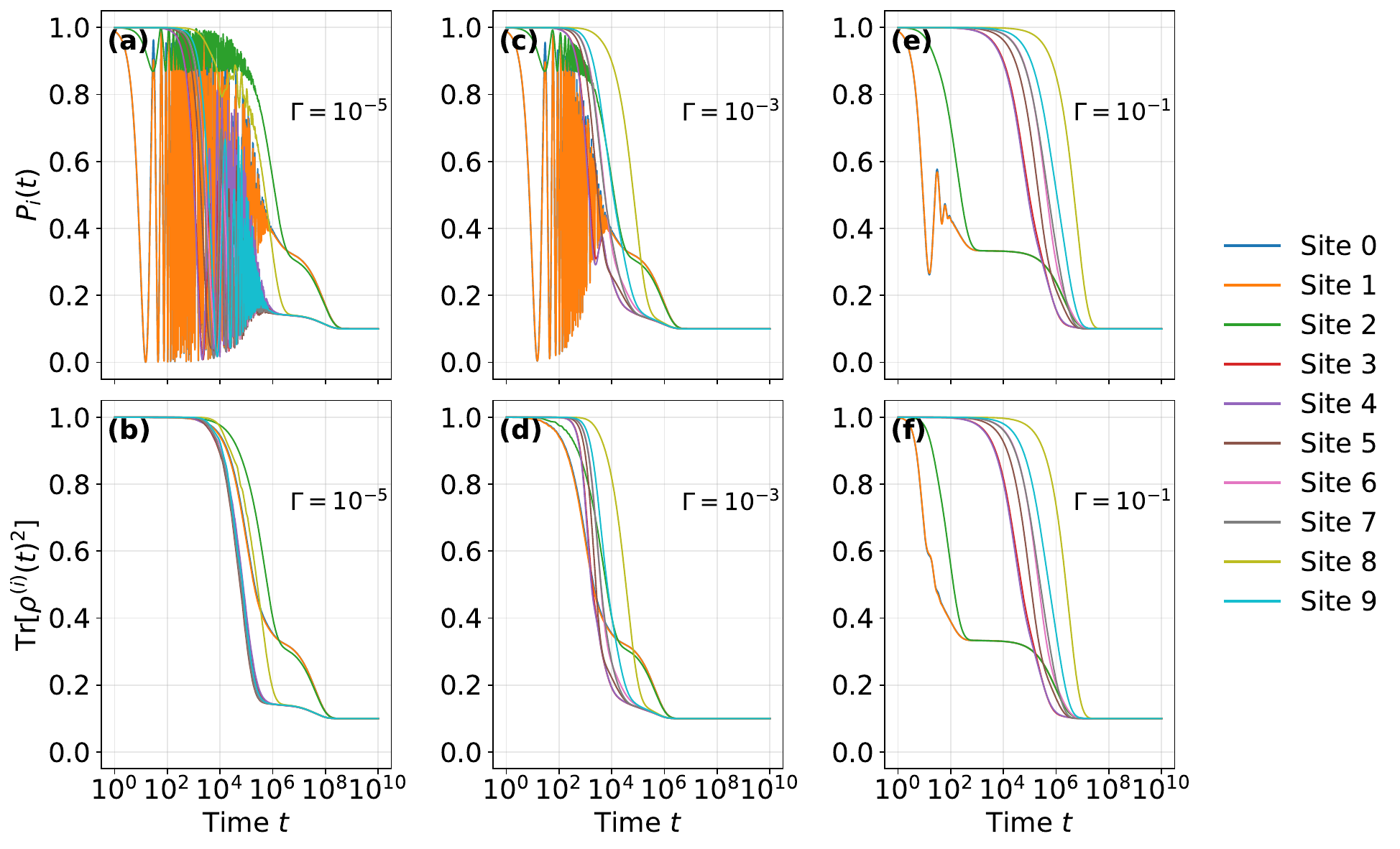}
\caption{
Dynamics of population and purity in Configuration A.
Top row: Population evolution $P_j(t)$. Bottom row: Purity.  Results are shown for three dephasing strengths: (a)-(b) $\Gamma=10^{-5}$; (c)-(d) $\Gamma=10^{-3}$; and (e)-(f) $\Gamma=10^{-1}$.
Each curve represents a simulation performed with a different initial excited site, $\rho(0) = |i\rangle\langle i|$. 
In the weak dephasing regime, the strongly hybridized cluster retains the excitation for parametrically longer times, following the $J^2/(\epsilon^2\Gamma)$ scaling predicted by the minimal model. 
The final state achieved at each case approaches the fully mixed state with uniform population  $1/N$ on each site.
}
\label{fig:geomA_dynamics}
\end{figure*}
\twocolumngrid

\subsection{Population dynamics}

The hierarchal relaxation process can be exposed by following the population dynamics. In Fig. \ref{fig:geomA_dynamics}, the system is examined when initialized in ten different initial conditions, with the excitation sitting at one of the sites 0 to 9. We increase the local dephasing $\Gamma$
(left to right) and present both the site populations as a function of time,
Fig. \ref{fig:geomA_dynamics}(a)-(c), and the state purity as it evolves, Fig. \ref{fig:geomA_dynamics}(d)-(f).

In the weak dephasing regime, shown in panels (a)–(b), initial oscillations are suppressed at times $t \approx \Gamma^{-1}$. For intermediate times, $1/\Gamma < t < J^2/(\Gamma \epsilon^2)$, the system undergoes a local, internal equilibration within a subset of sites, which effectively form a subnetwork.
For example, when the system is initialized at site $0$, the population and purity of its subnetwork in this intermediate regime approach $\sim 0.33$, reflecting local equilibration among the three neighboring sites ($0$, $1$, and $2$). The excitation remains confined within this subset for an extended period due to the ``triplet fortress effect,''  described in Sec.~\ref{sec: Minimal model analysis: Timescale Hierarchies} for the case of twin sites.
Similarly, for initial conditions on any of sites $4$ to $9$, the population and purity reach values of approximately $0.14$ within the same intermediate timescale, again reflecting local subnetwork equilibration. At later times, $t > J^2/(\Gamma \epsilon^2)$, the system fully equilibrates, with both population and purity approaching $1/10$, consistent with a completely-mixed state forming over all ten sites.
Similar behavior is observed in the intermediate dephasing regime, $\Gamma = 10^{-3}$, as shown in Fig.~\ref{fig:geomA_dynamics}(c)–(d).

In the strong dephasing regime, $\Gamma > J$, we show in panels (e)–(f) that if the system is initialized within the strongly-coupled group (sites 0, 1, and 2), the equilibration proceeds in two stages. First, there is an initial relaxation within the local group, during which the system remains confined for times $\Gamma/J^2 < t < \Gamma/\epsilon^2$. This is followed by a second stage, where complete equilibration with the rest of the ensemble occurs. 
In contrast, if the system is initialized with excitations in the larger group of spins (sites 3 to 9), the population of these sites decays on a single timescale, $\Gamma/\epsilon^2$.

In Appendix~\ref{AppA}, we present an alternative approach to analyzing the dynamics by examining the Liouvillian (dynamical map) in terms of its eigenvalues and eigenvectors. The eigenvalues encode the relaxation rates, or equivalently, the associated timescales, as well as the characteristic oscillation frequencies. The projections of the eigenvectors onto the site populations quantify the extent to which each site participates in the corresponding relaxation modes.

\begin{figure}[t!]
\centering
\includegraphics[width=1\linewidth]{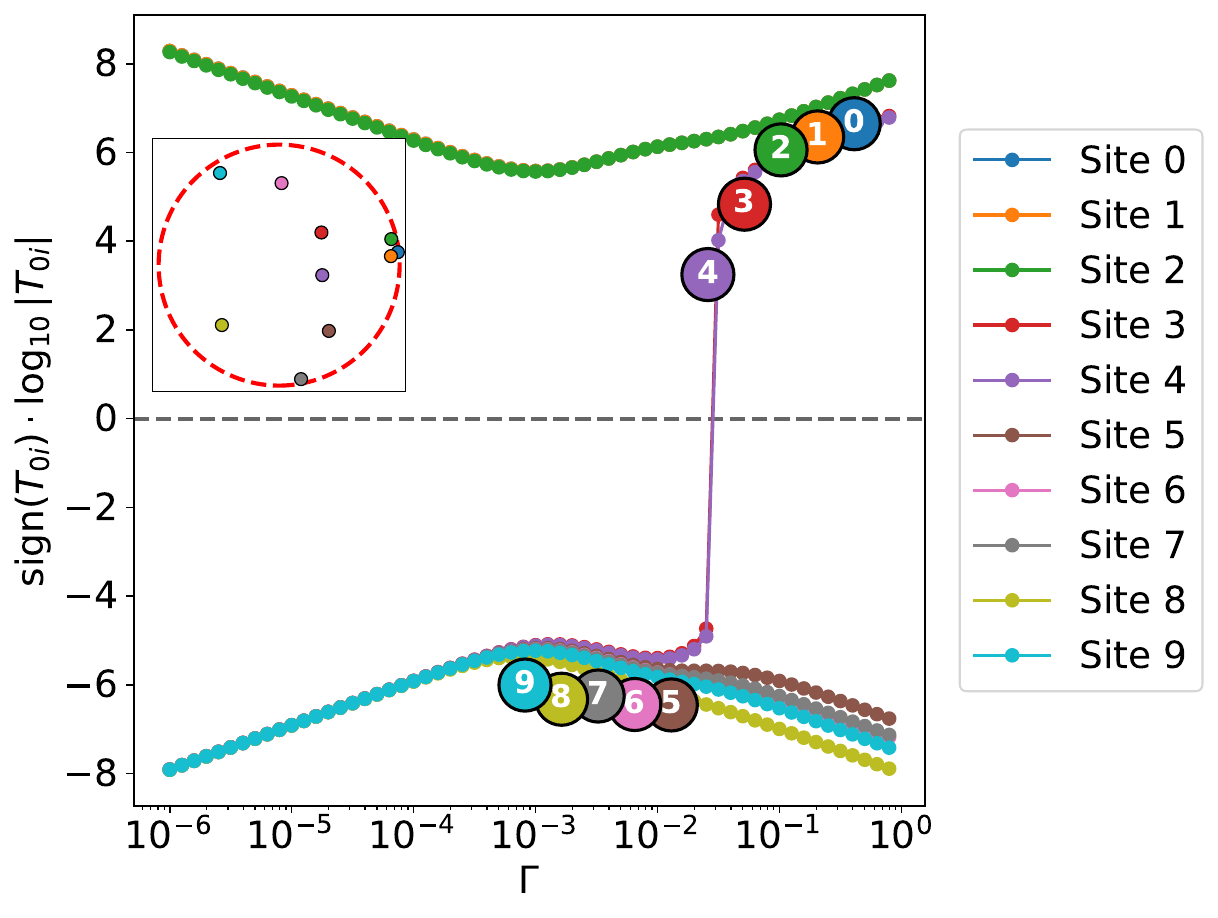}
\caption{
Configuration A: Integrated transfer measure $T_{0i}$ as a function of the dephasing strength $\Gamma$, with the system initialized at site $0$. Sites $0$--$2$ exhibit positive values, reflecting transient population accumulation within this strongly coupled cluster above equilibrium value. More distant sites show negative $T_{0i}$, reflecting that their transient population lies below their equilibrium value.
Sites $3$ and $4$, which bridge the two subnetworks, display a non-monotonic behavior, and undergo a sign change near $\Gamma \sim J$. This crossover signals the suppression of coherent hybridization in the 0-2 cluster, and identifies these sites as dynamical bottlenecks separating fast local equilibration from slower global relaxation.
For clarity, the inset displays Configuration A.
}
\label{fig:geom1T0i}
\end{figure}

\begin{figure}[t!]
\includegraphics[width=1\linewidth]{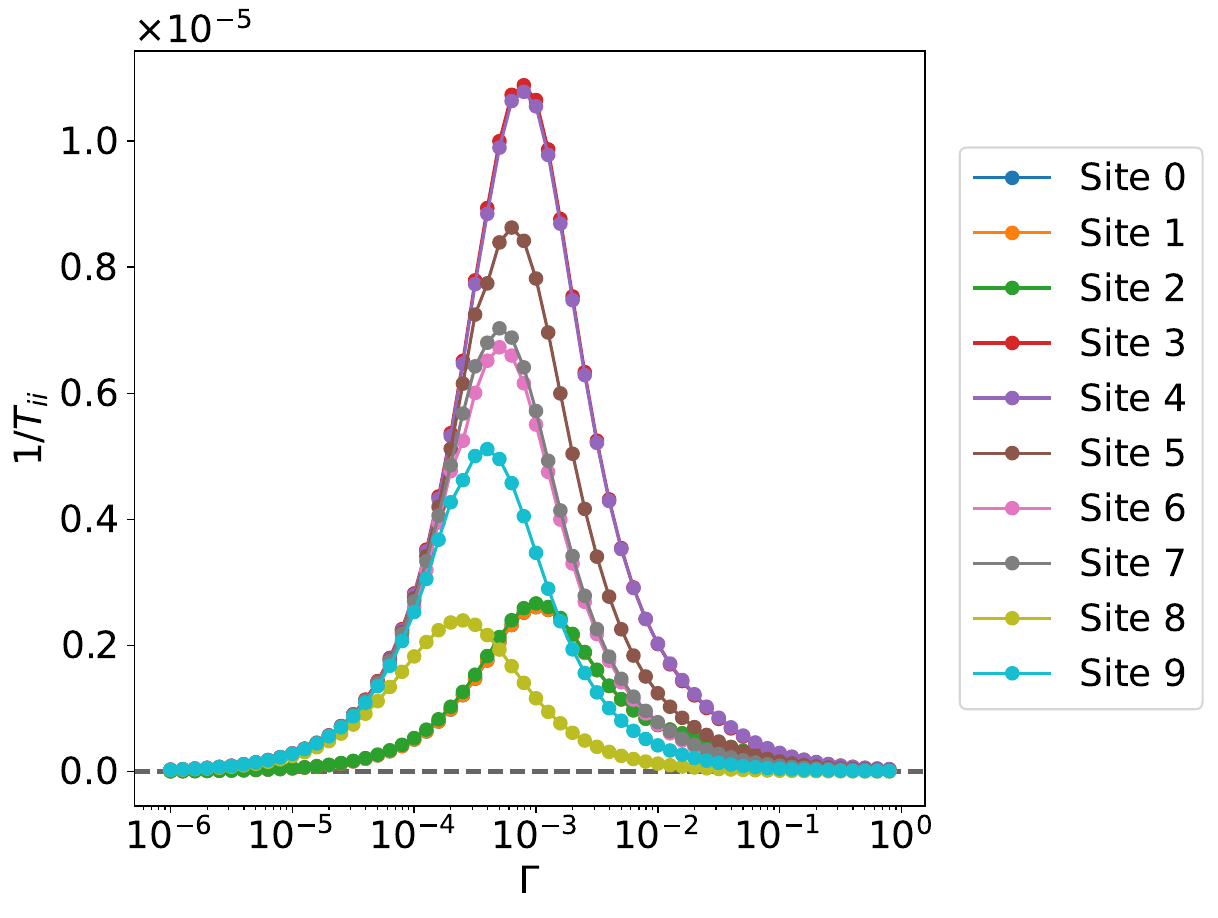}
\caption{
Configuration A: Inverse of the survival measure $T_{ii}^{-1}$, plotted as a function of the dephasing strength $\Gamma$ for all sites. 
The strongly coupled sites $0$ to $2$ exhibit a rightward shift of the minimum. 
}
\label{fig:geom1Tii}
\end{figure}

\begin{figure}[t!]
\includegraphics[width=1.0\linewidth]{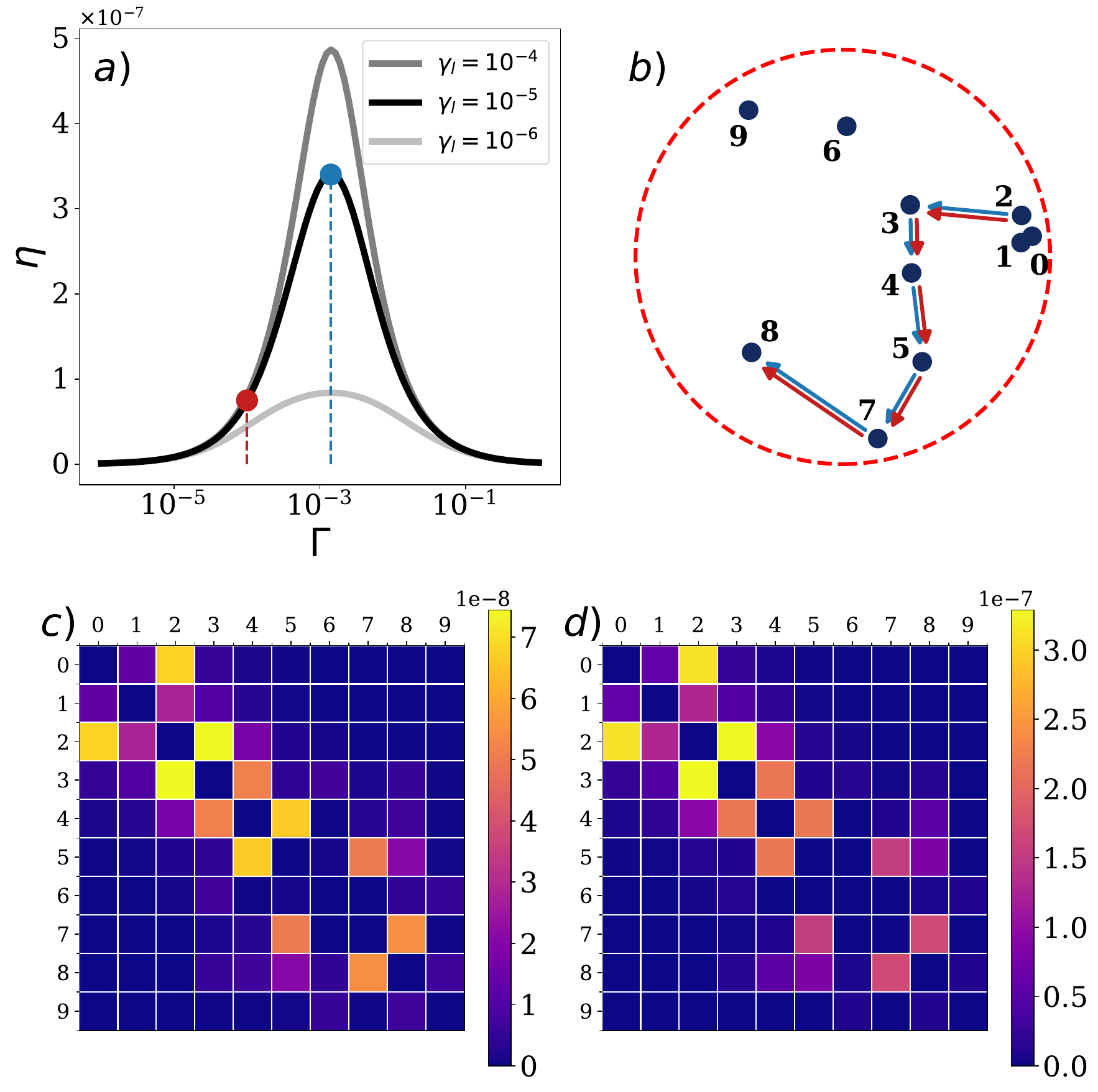}
\caption{
Nonequilibrium steady state in configuration A.
(a) Population flux $\eta=\gamma_l\bra{8}\hat{\rho}^\text{NESS}\ket{8}$ plotted as a function of dephasing strength $\Gamma$ at different injection/extraction rate constant $\gamma_l$. The injection (extraction) is located at site 0 (8) for the entire simulation. The red and blue vertical lines at $\Gamma=10^{-4}$ and $1.419\times 10^{-3}$ are selected to probe whether distinct transport channels emerge as $\Gamma$ is varied. 
(b) The dominant steady-state transport channel in configuration A is marked, inferred from the  absolute value of the probability current plots $|F_{mn}|=|2H_{nm}\Im \rho_{mn}^\text{NESS}|$ at (c) $(\Gamma=10^{-4}$  and (d) $1.419\times 10^{-3}$.      
The probability current between sites $m$ and $n$, $F_{mn}$, is determined by the imaginary part of the coherence and the tunneling energy. Large non-zero values of $|F_{mn}|$ mark the most dominant path from $0\rightarrow8$.}
\label{fig:NESS_A}
\end{figure}


\subsection{Integrated survival and transfer time measures}

We next demonstrate that the transfer-time measure can be used to reveal the \emph{connectivity} of the network, as well as the presence of embedded subnetworks. Complementing it, the inverse of the survival-time measure identifies the dephasing energy scale that is optimal for overcoming the ``fortress'' effect, thereby enhancing the extraction of the excitation from the initial site. As such, this measure provides direct access to the relevant energy scales that govern the dynamics.

We continue with Configuration A, presented in Fig. \ref{fig:geom1}.
Fig.~\ref{fig:geom1T0i} presents the transfer-time measure $T_{0i}$ for this geometry. It probes how the population initially prepared on site 0 is redistributed across the network prior to equilibration. According to Eq.~(\ref{eq:Tij}), the first index denotes the preparation site, while the second index labels the observation site. 
A positive value of $T_{0i}$ indicates accumulation of the transient population above the steady-state value, while a negative value signifies depletion relative to equilibrium. The magnitude of $T_{0i}$ characterizes the integrated timescale over which site $i$ participates in the relaxation dynamics.

In the weak dephasing regime, sites $0$, $1$, and $2$ exhibit nearly overlapping curves (appearing at the top), identifying them as a strongly hybridized subsystem. The population rapidly redistributes within this cluster before it reaches the rest of the network. The remaining sites ($3$ to $9$) form a second group, corresponding to the lower set of curves.

Sites $3$ and $4$ interpolate between the two structures and act as dynamical bridges. While for small $\Gamma$ they appear as part of the large group, as $\Gamma$ increases and becomes comparable to the internal tunneling scale $J\approx 0.01$, these sites undergo a change in their dynamical role and begin to align with the cluster behavior of sites $0$, $1$ and $2$. This crossover reflects the suppression of coherent intra-cluster hybridization in the 0,1,2 subnetwork, and the activation of inter-cluster transfer.

The transfer-time measure, $T_{0i}$, thus enables the identification of distinct clusters within the configuration, as well as sites that mediate connectivity between them. Furthermore, it provides access to the characteristic energy scale  of the strongly coupled cluster, as reflected by the value of $\Gamma$ at which the bridging behavior emerges.

We now turn to the analysis of the survival time measure, $T_{ii}$, shown in Fig.~\ref{fig:geom1Tii}, defined as the time-integrated population {\it remaining} on the initially excited site $i$, prior to equilibration, see Sec.~\ref{sec: Physical Setup, Model and dynamical Measures}. We plot the \emph{inverse} of the survival-time measure so that larger values correspond to faster relaxation from the initialized site. Each curve corresponds to a different initial condition and thus reflects the tendency of the system to transfer population away from the corresponding site.

We make the following observations: 

(i) Sites $0$, $1$, and $2$ (overlapping under the curve of site 2) again display nearly identical behavior, confirming that they function as a collective subsystem. 
For small dephasing below $\Gamma^*$, where $1/T_{ii}$ is maximal, coherent hybridization within the cluster gives rise to localized dynamics. Relaxation out of the cluster is then governed by a hierarchical leakage timescale $t_{\rm tr} \sim \frac{J^2}{\epsilon^2 \Gamma}$. Increasing dephasing enhances inter-cluster transfer, until dephasing becomes comparable to tunneling in the tight cluster, at which point the excitation escapes more efficiently and populates the rest of the spin network.

(ii) Sites $3$ to $9$ belonging to the larger subnetwork
are similarly governed by competing effects. For very small $\Gamma$, cluster-induced energy detuning suppresses transfer from this group.  For very large $\Gamma$, global Zeno suppression reduces effective tunneling. 
Optimal relaxation therefore occurs in an intermediate regime 
where dephasing is strong enough to overcome cluster protection yet not large enough to suppress transport throughout the network.  

(iii) The shift in the position of the maximum between the two groups ($0$ to $2$ and $3$ to $9$) reflects the existence of distinct geometric coupling scales within the same model. 

%

\subsection{Nonequilibrium steady state flux analysis}
\label{sec: Steady state flux analysis config A}

The system can be analyzed in a complementary fashion through its nonequilibrium steady state (NESS), by introducing injection and extraction sites. The resulting population flux provides a probe of the dephasing scale $\Gamma$ required to overcome the ``fortress'' effect and optimize transport, in close analogy with the survival-time measure. More generally, NESS at varying dephasing strengths reveals the dominant transport pathways under these conditions.

As described in Sect.~\ref{subsec: Measures for relaxation and transfer processes}, the nonequilibrium steady-state for the flux analysis is created by enforcing a constant injection (extraction) flux at site $n$ $(m)$. This is achieved by adding a jump operator $\hat{L}_l=\sqrt{\gamma_l}\ket{n}\bra{m}$ to the Lindblad QME describing the model under local dephasing noise. To solve the equation, we first write the QME in the doubled Hilbert space where the density matrix is written in a vectorized form:
\begin{equation}
   \frac{d}{dt}|\hat{\rho}(t)\rangle\!\rangle = \hat{\mathcal{L}}|\hat{\rho}(t)\rangle\!\rangle.
\end{equation}
Here, $|\hat{\rho}(t)\rangle\!\rangle$ is the vectorized density matrix and $\hat{\mathcal{L}}$ is the Liouvillian superoperator. The NESS satisfies
\begin{equation}
    \hat{\mathcal{L}}|\hat{\rho}^\text{NESS}\rangle\!\rangle = 0,
\end{equation}
as the state is stationary in time. Next, we impose the trace-preserving condition of the density matrix by replacing the row corresponding to the population of the injection site $n$ 
of $\hat{\mathcal{L}}$ with the vectorized identity matrix. We denote this superoperator, with the trace-preserving condition imposed, as $\tilde{\mathcal{L}}$. Finally, we solve
\begin{equation}
    \tilde{\mathcal{L}}|\hat{\rho}^\text{NESS}\rangle\!\rangle = |1_n\rangle\!\rangle,
\end{equation}
where $|1_n\rangle\!\rangle$ corresponds to a vectorized form of $\ket{n}\bra{n}$ (That is, a vector of zeros except for the element corresponding to $\rho_{nn}=1$). The solution is then obtained by $|\hat{\rho}_\text{NESS}\rangle\!\rangle = \tilde{\mathcal{L}}^{-1}|1_n\rangle\!\rangle$. 

Once the NESS is found, we reshape $|\hat{\rho}^\text{NESS}\rangle\!\rangle$ back into its matrix form and define the population flux at the extraction site $m$ as 
\begin{equation}
    \eta = \gamma_l \bra{m} \hat{\rho}^\text{NESS}\ket{m}.
\end{equation}
Results for Configuration A are shown in Fig.~\ref{fig:NESS_A}.
In this scenario, population is injected into site $0$ and extracted from site $8$. Specifically, in Fig.~\ref{fig:NESS_A}(a), the population flux at site $8$ is plotted as a function of local dephasing rate $\Gamma$ for three different values of the extraction rate constant $\gamma_l\in[10^{-4},10^{-5},10^{-6}]$. Qualitatively similar behavior is found in all cases: The flux is optimized around \(\Gamma \simeq 10^{-3}\), where it is maximized relative to other values of \(\Gamma\) while other parameters are held fixed. 

In Fig.~\ref{fig:NESS_A}(c) and (d), we further illustrate the excitation transfer pathways by plotting the probability current on each site. We consider two representative values of $\Gamma\in[10^{-4},1.419\times 10^{-3}]$, denoted by red and blue points in Fig.~\ref{fig:NESS_A}(a), respectively, for $\gamma_l=10^{-5}$.
The probability current between sites $m$ and $n$ is determined by the imaginary part of the coherence, weighted by the tunneling coupling \cite{Liu_2019}, 
\bea
F_{mn}=2H_{nm}\Im \rho_{mn}^\text{NESS},
\eea
with $\hat H$ as the Hamiltonian of the network, Eq.~\eqref{eq:ham}.

The magnitude of the probability current, $|F_{mn}|$, reveals the dominant transport pathway connecting site 0 to site 8. For example, Fig.~\ref{fig:NESS_A}(c) shows pronounced probability flux along the sequence of links
$0 \leftrightarrow 2 \leftrightarrow 3 \leftrightarrow 4 \leftrightarrow 5 \leftrightarrow 7 \leftrightarrow 8$.
This sequence therefore defines the dominant pathway for population transfer from site $0$ to site $8$ at $\Gamma=10^{-4}$, as illustrated schematically in Fig.~\ref{fig:NESS_A}(b). A corresponding analysis at larger dephasing, near the peak transport regime [Fig.~\ref{fig:NESS_A}(d)], reveals essentially the same transport route. Overall, similar current patterns emerge across different dephasing strengths, indicating the presence of a robust dominant transport pathway highlighted in Fig.~\ref{fig:NESS_A}(b).

\begin{figure}[htpb]
\includegraphics[width=0.85\linewidth]{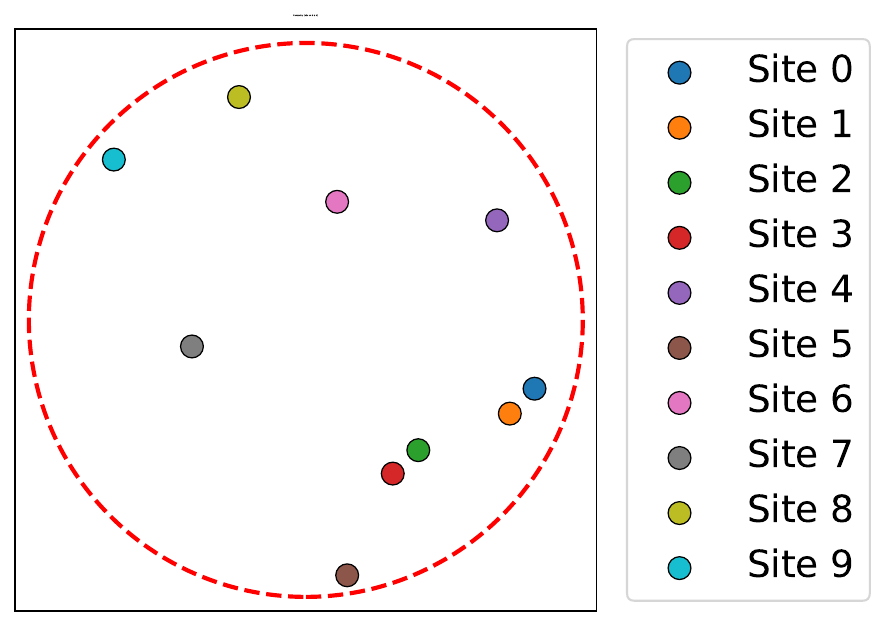}
\caption{Configuration B.
Representative realization of a ten-site spin ensemble within a circular region. Sites are labeled according to the numbering used throughout the dynamical analysis. }
\label{fig:geom2}
\end{figure}

\begin{figure*}[htbp]
\centering
\begin{subfigure}{0.48\textwidth}
    \centering
    \includegraphics[width=\linewidth]{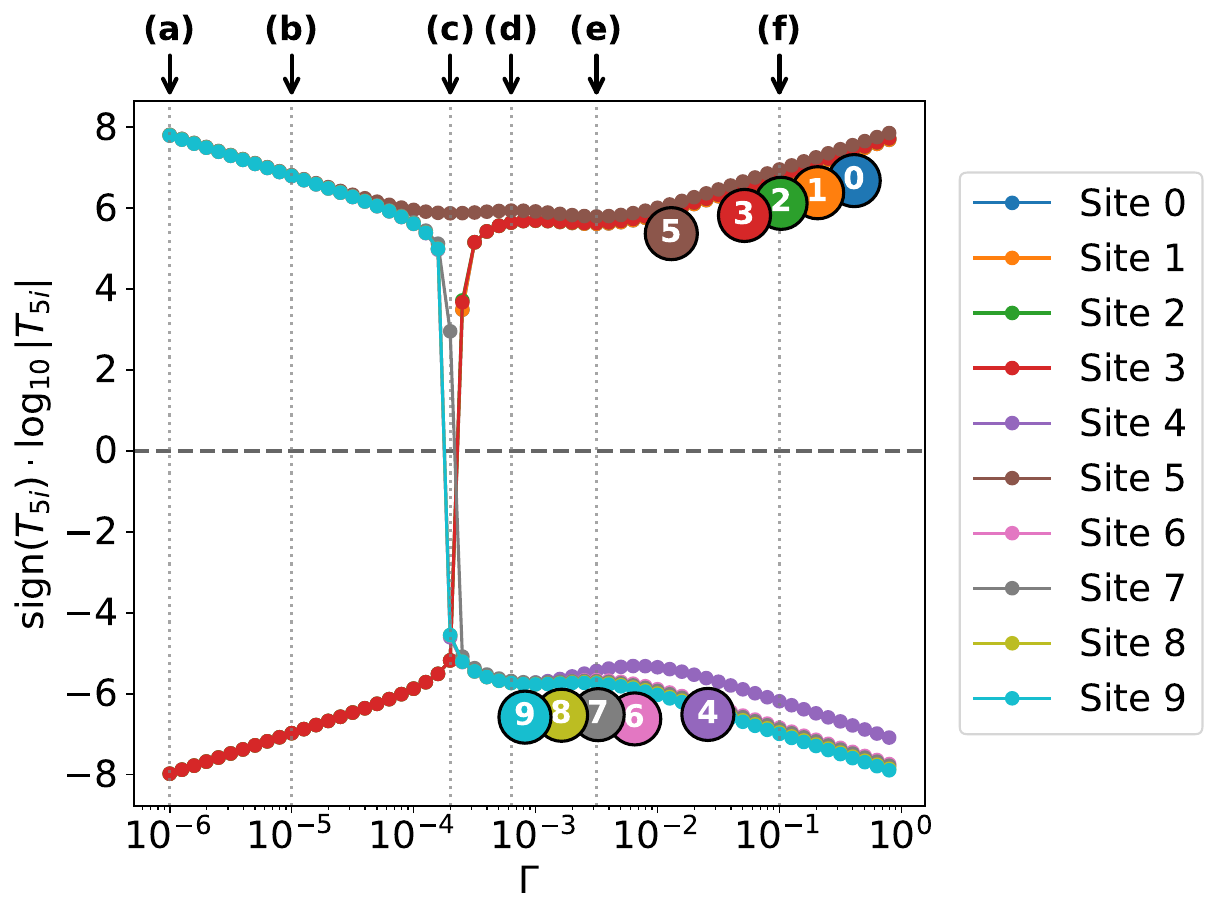}
\end{subfigure}
\hfill
\begin{subfigure}{0.48\textwidth}
    \centering
    \includegraphics[width=\linewidth]{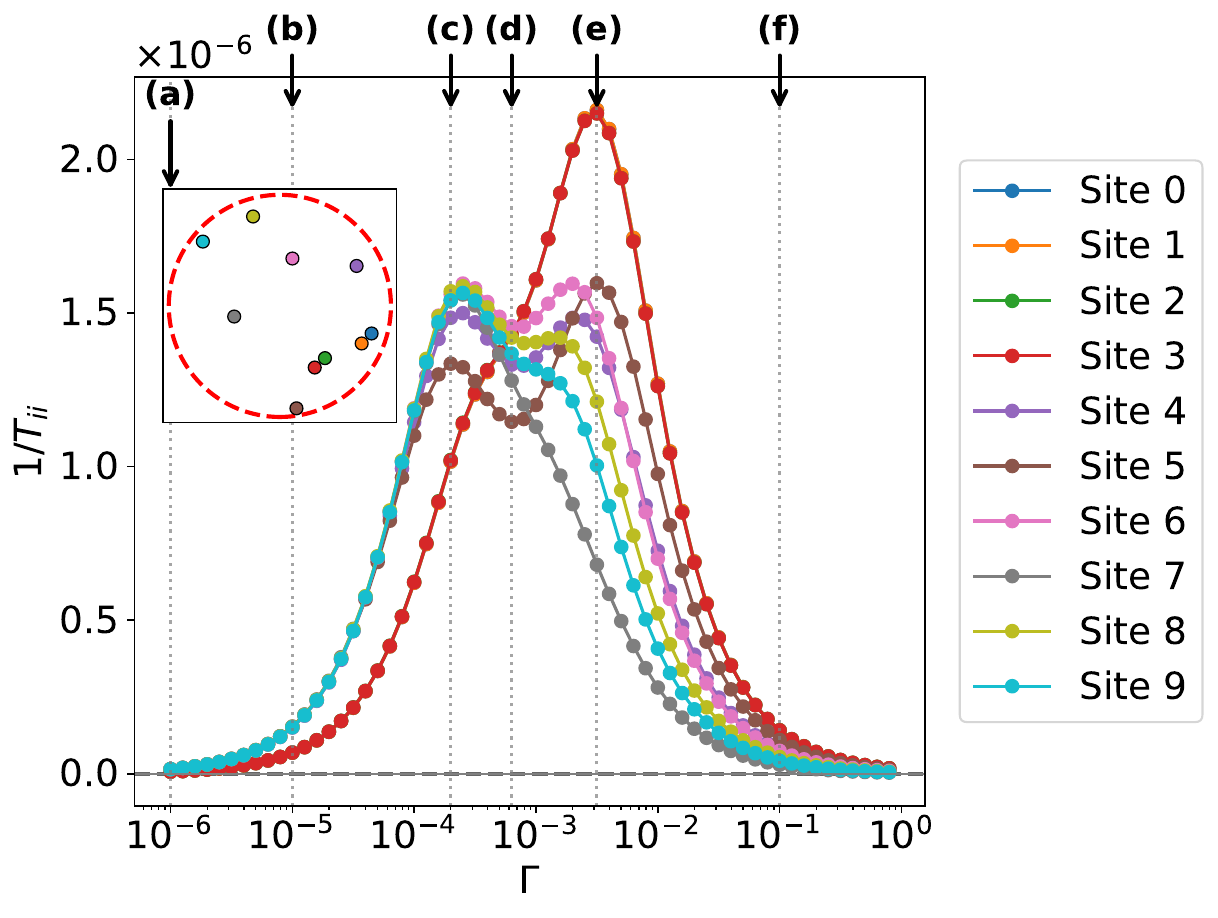}
\end{subfigure}
\caption{
Configuration B: Left panel shows the transfer measures $T_{5i}$ and right panel the survival measures $T^{-1}_{ii}$ as a function of $\Gamma$. The arrows mark the six dephasing values corresponding to the dynamical regimes in Fig. \ref{fig:geom2_dyn}. A clear double peak emerges from the competition between two transport pathways. As $\Gamma$ increases, dominance switches from transfer through site $9$ to transfer through sites $0$--$3$, reflected by the exchange of sign and magnitude in $T_{5i}$. For small $\Gamma$, sites $4$--$9$ all exhibit positive $T_{5i}$, indicating that the dominant transport pathways initially involve the left hand side of the network.
}

\label{fig:geom2_T}
\end{figure*}

\begin{figure*}[htbp]
\centering
\includegraphics[width=\textwidth]{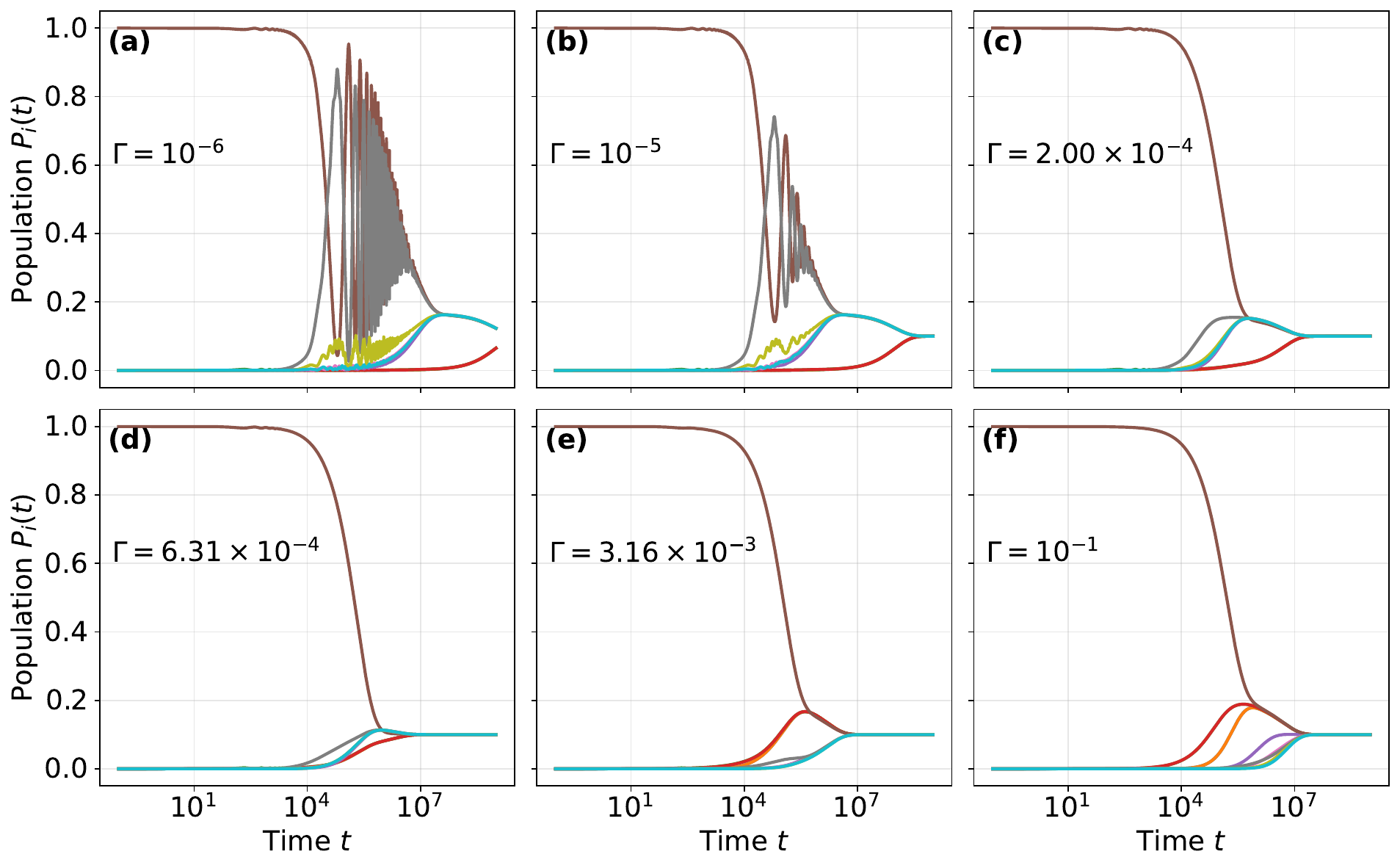}
\caption{
Configuration B: Population dynamics at each site for six representative dephasing strengths spanning the weak, intermediate-crossover, and strong regimes ($\Gamma = 10^{-6}, 10^{-5}, 2\times10^{-4}, 6.31\times10^{-4}, 3.16\times10^{-3}, 10^{-1}$), with the excitation initialized at site $5$. 
Color scheme is the same as in Fig. \ref{fig:geom2_T}.
At weak and intermediate dephasing, population transfer proceeds predominantly through site $7$ before spreading to the rest of the cluster. 
As $\Gamma$ increases and becomes comparable to the strong tunneling scale, hybridization within the strongly coupled subcluster is suppressed, activating an alternative equilibration pathway through sites $0$–$3$. 
This pathway switching reshapes the relaxation dynamics and underlies the emergence of the double peak behavior observed in the survival and steady state measures.
}
\label{fig:geom2_dyn}
\end{figure*}

\section{Configuration B: Competing relaxation pathways} 
\label{sec:ConfB}

As a second example of relaxation dynamics in heterogeneous systems, we consider another random configuration, shown in Fig.~\ref{fig:geom2}. This configuration enables us to investigate the presence of competing equilibration pathways and the environmental conditions under which different pathways are favored. 

The following analysis focuses on configurations in which the excitation is initially prepared on site 5, located at the bottom of the network. The resulting relaxation dynamics and transport pathways are characterized using several complementary measures: the time-dependent site populations, $P_i(t)$; the transfer measure, $T_{5i}$; and the nonequilibrium steady-state flux, with site 5 serving as the injection site and the opposite site 8 (top) acting as the extraction site. In addition, the inverse survival-time measure, $T_{ii}^{-1}$, is reported for each site.

The central dynamical feature of Configuration ~B, as evidenced by Figs.~\ref{fig:geom2_T}, \ref{fig:geom2_dyn}, and \ref{fig:NESS_B}, is the emergence of two distinct relaxation pathways that dominate under different dephasing regimes. 
Starting from site $5$, in the weak dephasing regime, the excitation predominantly relaxes to the rest of the network via site $7$, which serves as the primary mediator connecting the initial site to the rest of the network. In contrast, at large $\Gamma$, when the dephasing becomes comparable to the strong internal coupling scale of the subcluster formed by sites $0$ to $3$, coherent hybridization within this subcluster is suppressed. 
This, in turn, activates an alternative relaxation pathway, initiated by transfer from site $5$ into the strongly coupled region comprising sites $0$ to $3$. 
We now elaborate on these findings.

\subsection{Integrated survival and transfer time measures}

Fig.~\ref{fig:geom2_T} summarizes the integrated transfer diagnostics for Configuration B. We focus on initialization at the site $5$ and follow the transfer measure $T_{5i}$. We also study the inverse survival measure at each site, $T_{ii}^{-1}$. These metrics are presented as functions of the dephasing strength $\Gamma$. To connect these measures to the time domain dynamics, in Fig.~\ref{fig:geom2_T} we mark six representative values of the dephasing strength,
$\Gamma = 10^{-6},\,10^{-5},\,2\times 10^{-4},\,6.31\times 10^{-4},\,3.16\times 10^{-3},\,10^{-1}$,
denoted by (a) through (f), with the corresponding dynamics presented in Fig.~\ref{fig:geom2_dyn}.
Points (a), (b), and (f) probe the two asymptotic weak dephasing regimes and a strong dephasing reference point, respectively, while points (c) and (e) lie near the two local maxima in $T_{ii}^{-1}$. Point (d) lies near the local minimum between them.

The transfer measure $T_{5i}$ shown in Fig.~\ref{fig:geom2_T} (left) provides a clear view of defined subclusters in the network at different values of the dephasing rate. At weak dephasing, site $5$ forms a subcluster with sites $7$ and $9$ (top left), with which it equilibrates early in the dynamics.
As $\Gamma$ is tuned through a crossover, a collective sign and magnitude redistribution occurs across sites, indicating that the integrated population weight shifts from the pathway associated with sites $7$ and $9$ to a pathway involving the strongly coupled subcluster of sites $3$ to $0$. 


Competition in pathways is also reflected in the double peak structure of $T_{ii}^{-1}$ in Fig.~\ref{fig:geom2_T} (right).
The first peak at weak dephasing corresponds to a regime in which moderate dephasing enhances transfer along the pathway mediated by site $7$. 
The second peak appears at higher $\Gamma$, when dephasing has sufficiently weakened the coherent protection of the strongly coupled subcluster, thereby enabling efficient transport through sites $0$ to $3$.  The local minimum between the peaks reflects the crossover region in which neither pathway is yet dominant.

Presented population dynamics, Fig.~\ref{fig:geom2_dyn} substantiate these results clearly showing the transition between the two relaxation pathways when increasing $\Gamma$. The curves correspond to the population building at each site of the ensemble, with the initial condition defined at site $5$.
In weak dephasing, site $7$ initially acts as a mediator to the overall relaxation dynamics; see Fig.~\ref{fig:geom2_dyn}(a)-(c). However, as the dephasing strength is increased, and once $\Gamma$ reaches the range of the large hybridization between sites $0$-$3$, they serve as the gateway for the relaxation dynamics, see Fig.~\ref{fig:geom2_dyn}(e)-(f).

Thus, Configuration B differs qualitatively from Configuration A. 
In Configuration A, relaxation is governed by a single hierarchical relaxation structure, which shifts the optimal dephasing condition but does not generate multiple maxima. In contrast, Configuration B supports two competing transport pathways controlled by distinct coupling scales. 
Because each pathway becomes optimal in a different dephasing window, the system exhibits two separate optimal regimes and therefore a double peak in the survival and transport  measures. 

The competition between transport pathways, and the resulting ``double peak" structure observed in the survival measure, directly addresses the open question of this observed feature in previous works \cite{Eriktwin23}. The identification of this behavior with underlying network features supporting competing transport pathways, activated under different dephasing regimes, is a central result of this work. This effect is further investigated below using the NESS framework,  together with an analysis of its robustness from both statistical and ensemble perspectives.
\subsection{Nonequilibrium steady state flux analysis}

Studies of environment-assisted quantum transport have predominantly focused on quasi-one-dimensional chains and specific models of light-harvesting complexes
\cite{Alan08,Plenio08,Plenio09,Plenio10,Plenio12,Cao13,Plenio21,mohseni2014energy,trautmann2018,ZH2018,cygorek2022,kurt2023,Cao09,Maggie25,Rebentrost2009,Coates2021,Coates2023,Lawrence2026arxiv}.
In such systems, transport efficiency is typically optimized within a specific range of environmental coupling strengths. In contrast, Ref.~\cite{Eriktwin23} recently demonstrated that more complex transport networks can exhibit multiple optimal peaks in their steady-state transport efficiency. However, the microscopic structure-function relationship underlying this behavior remained unresolved, in particular the connection between network geometry and the emergence of single or multiple efficiency peaks as a function of the dephasing rate, $\Gamma$.

Configuration A supports a single-peak flux-dephasing curve; see Fig.~\ref{fig:NESS_A}(a), indicating that transport is optimized within a single range of the dephasing rate $\Gamma$.  
In contrast, Configuration B, as suggested by the transient analysis above, points to qualitatively different behavior. We proceed to investigate the NESS transport properties of this setup, and rationalize the emergence of a double-peak structure. Based on the present analysis, this phenomenon is attributed to the competition between two transport pathways.

\begin{figure}[t!]
\includegraphics[width=1.0\linewidth]{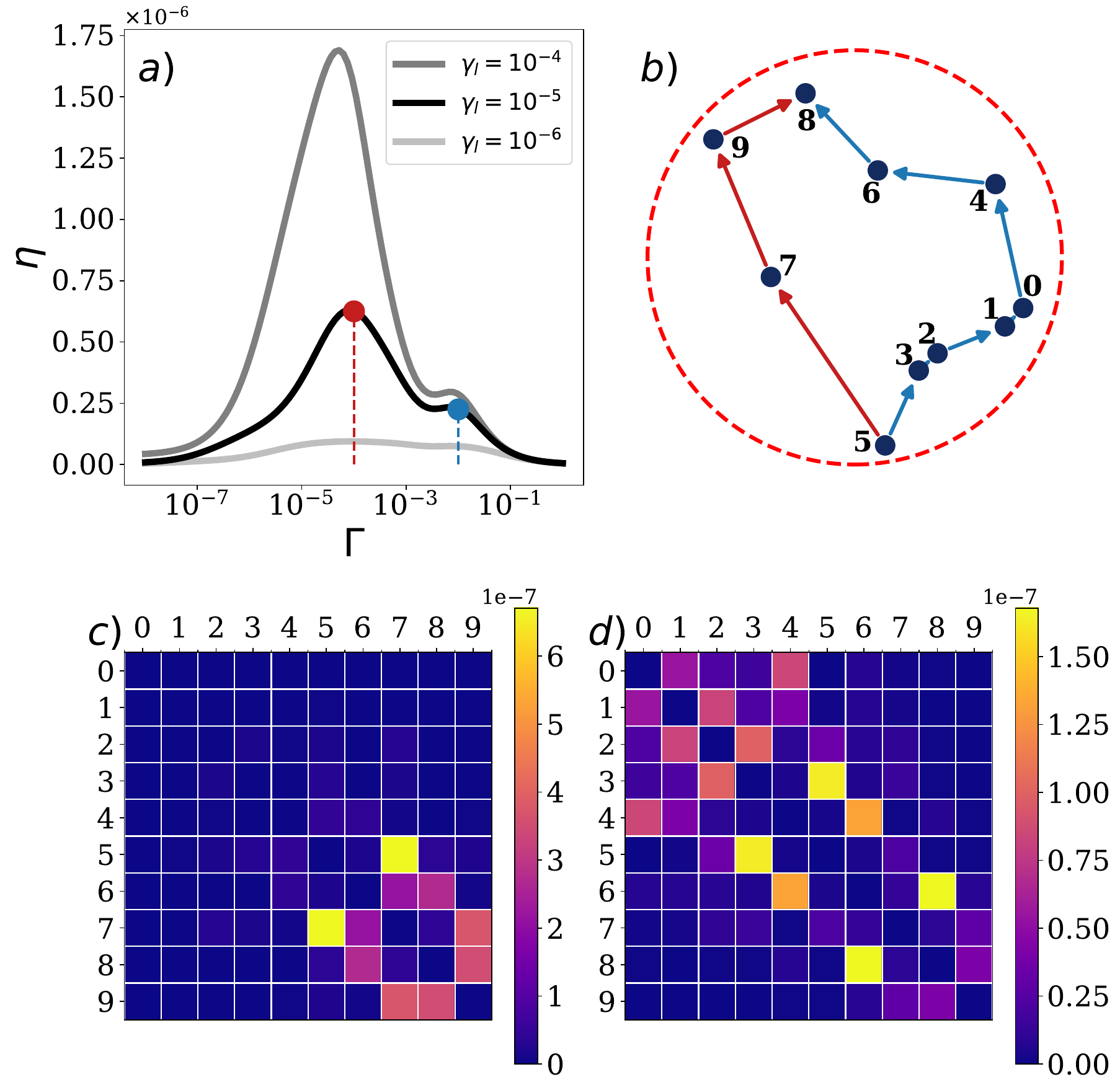}
\caption{
Nonequilibrium steady state in configuration B.
(a) Population flux $\eta=\gamma_l\bra{8}\hat{\rho}^\text{NESS}\ket{8}$ as a function of dephasing strength $\Gamma$ using different injection/extraction rate $\gamma_l$. The injection (extraction) is located at site $5$ ($8$) for the entire simulation. The red and blue vertical lines at $\Gamma=10^{-4}$ and $10^{-2}$ correspond to the approximate $\Gamma$ values where the two peaks stand for $\gamma_l=10^{-5}$. (b) The two peaks correspond to the dominant steady-state transport channels depending on $\Gamma$. This can be inferred from the absolute value of the probability current plots $|F_{mn}|=|2H_{nm}\Im \rho_{mn}^\text{NESS}|$ at (c) $\Gamma=10^{-4}$ and (d) $\Gamma=10^{-2}$.  The probability current between sites $m$ and $n$, $F_{mn}$, is determined by the imaginary part of the coherence and the tunneling energy. The magnitude of the values of $|F_{mn}|$ indicate on the most likely path from $5\rightarrow8$.}
\label{fig:NESS_B}
\end{figure}

We perform the NESS flux analysis as described in Sec.~\ref{sec: Steady state flux analysis config A}.
We inject the population into site $5$ and extract it from site $8$ (see Fig.~\ref{fig:geom2}). In Fig.~\ref{fig:NESS_B}(a), we plot the population flux at site $8$ as a function of the dephasing rate $\Gamma$ for three different values of the leakage rate $\gamma_l\in[10^{-4},10^{-5},10^{-5}]$ to demonstrate  robustness in trends. Unlike Configuration A, the NESS flux demonstrates a two-peak structure. We argue next that these two peaks indicate the presence of two competing excitation transfer mechanisms, see Fig.~\ref{fig:NESS_B}(b).
Specifically, there are two local maxima, one around $\Gamma=10^{-4}$ (red) and the other at $\Gamma=10^{-2}$ (blue), marked for $\gamma_l=10^{-5}$.  

Under the small value of $\Gamma$, the probability current plot in Fig.~\ref{fig:NESS_B}(c) shows that an excitation starting at site $5$ predominantly bypasses the tightly coupled sites $2$–$3$ and $1$–$0$, transferring population through sites
$5\to7\to9\to8$. In contrast, under large dephasing, Fig.~\ref{fig:NESS_B}(d) indicates that the dominant transfer pathway involves populating the two ``twin fortresses”  (2-3 and 1-0) since $\Gamma$ is large enough to compete with their internal tunneling coupling. 
In this case, the excitation travels from $5\to 3\to 2 \to 1 \to 0\to 4\to 6 \to 8$ as seen in the probability flux map.
In Fig.~\ref{fig:NESS_B}(b), we depict these two distinct pathways. 

Despite being a steady-state analysis, the NESS population flux qualitatively agrees with the dynamical simulations and measures presented in Figs.~\ref{fig:geom2_T} and \ref{fig:geom2_dyn} in predicting two distinct relaxation pathways, thus a two-peak behavior.  

\section{Statistical Signatures of Hierarchical Spin Baths}
\label{sec:hierarchical_statistics}

So far, we have focused on two representative configurations that highlight different aspects of the problem. In a random two-dimensional geometry, the distribution of coupling strengths is broad. A key outcome is that when sites are in close proximity such that their mutual coupling is strong compared to other couplings and to the dephasing rate, they effectively form a subcluster and undergo rapid internal equilibration (the ``twin fortress'' concept described in Sec.~\ref{sec: Minimal model analysis: Timescale Hierarchies}). Individual sites act as links between such clusters, facilitating equilibration across the full system.

Having established this behavior through case studies, we now aim to \emph{systematically} investigate the hierarchical relaxation dynamics that emerges when strongly connected subclusters (as small as two sites) are present.


\begin{figure}[t!]
\includegraphics[width=0.7\linewidth]{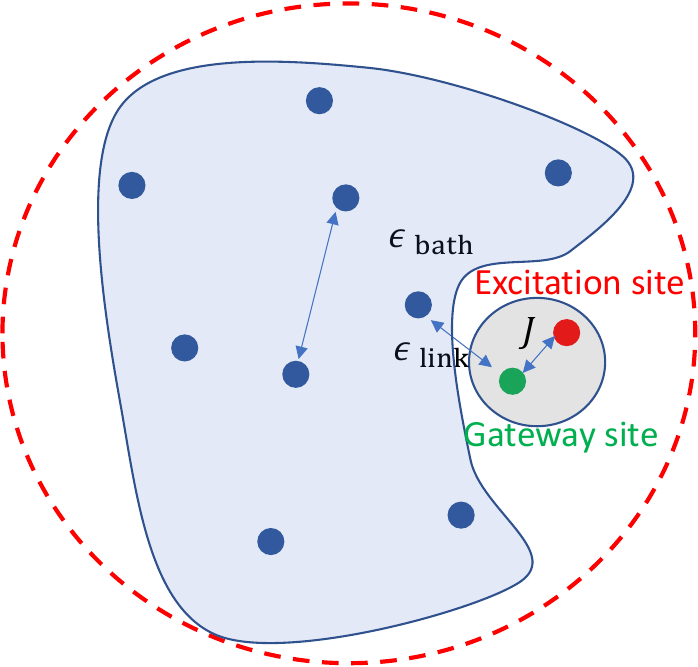}
\caption{Analysis of hierarchical relaxation from a particular site, denoted by ``excitation site", to the rest of the network.
The excitation site  is placed next to a ``gateway'' site, with a coupling strength $J$ that we artificially control. The gateway site connects the small cluster (excitation + gateway site) to the remainder of the system through a characteristic coupling parameter $\epsilon_{\mathrm{link}}$. The characteristic coupling within the bath network is denoted by $\epsilon_{\mathrm{bath}}$. In typical settings, $\epsilon_{\mathrm{bath}}/\epsilon_{\mathrm{link}} = 0.1\text{--}0.01$.
}
\label{fig:link}
\end{figure}

\begin{figure}[htbp]
\centering
\includegraphics[width=1\linewidth]{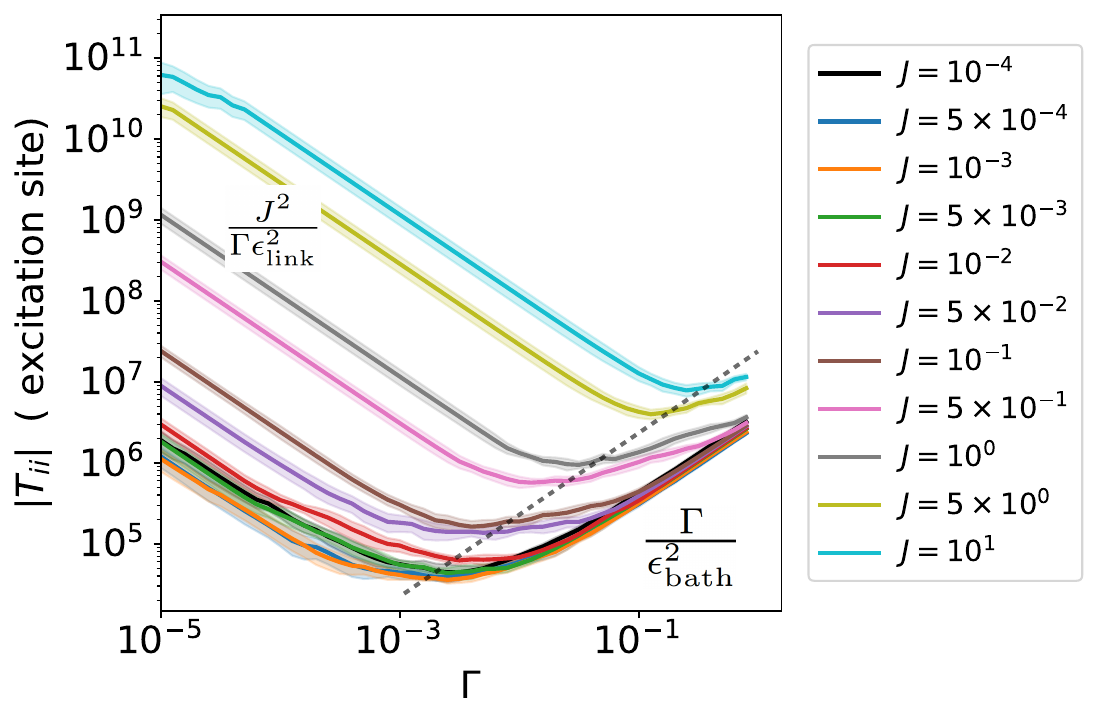}
\caption{
Integrated survival time measure from the excitation site, plotted against  $\Gamma$ for increasing $J$ value between the excitation site and the gateway site. 
The location of the minimum shifts to higher value in accord with $J$, as the hierarchy between the strong link and the rest of the bath develops. }
\label{fig:Tii_vs_gamma}
\end{figure}

\begin{figure*}[htbp]
    \centering
    \includegraphics[width=1\linewidth]{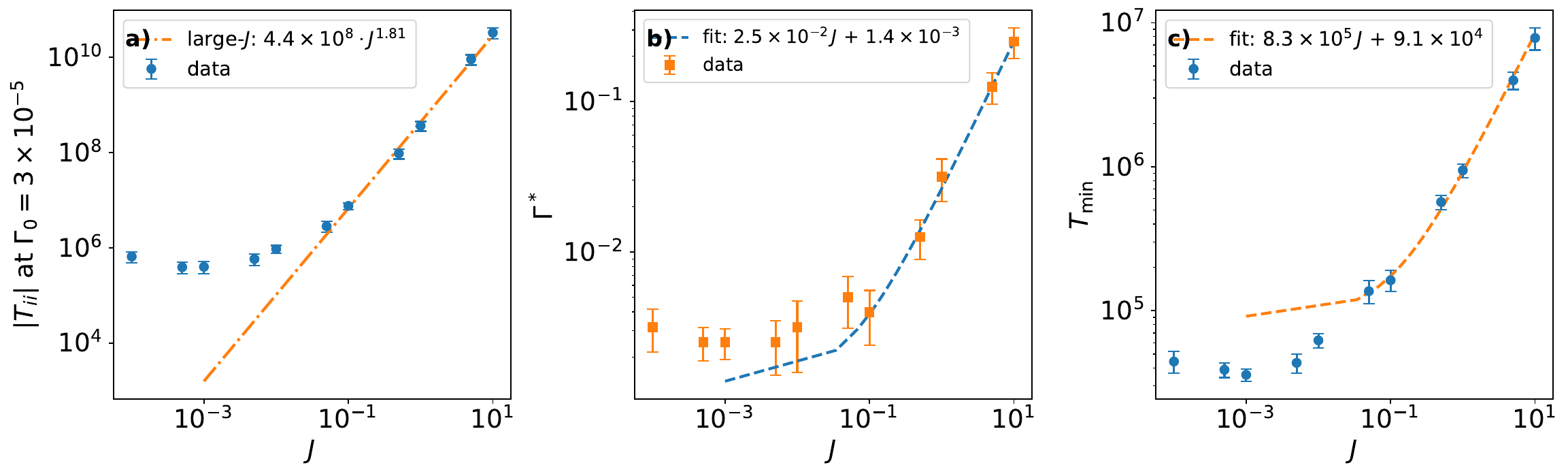}
    \caption{
Analysis of Fig.~\ref{fig:Tii_vs_gamma}. 
(a) The survival-time measure $T_{ii}$ at a fixed $\Gamma = 3.2 \times 10^{-5}$, plotted as a function of $J$, exhibits an approximate quadratic dependence. 
(b) The dephasing strength $\Gamma^{\ast}$ that minimizes the survival time, shown as a function of $J$, displays a linear scaling, $\Gamma^{\ast} \propto J$. 
(c) The minimal survival time, $T_{\min}$, plotted as a function of $J$, shows a linear scaling with $J$.
}
\label{fig:combined_fits}
\end{figure*}

\subsection{Generalized model}

In Sec.~\ref{sec: Minimal model analysis: Timescale Hierarchies}, a minimal three-site model with {\it two} distinct tunneling energy scales was introduced. In the weak-dephasing regime, where $\Gamma$ constitutes the smallest energy scale in the problem, a strongly hybridized pair of sites with coupling strength $J$ was shown to relax into the rest of the network on a parametrically long time scale, scaled as $J^2/(\Gamma \epsilon^2)$. Here, $\epsilon$ denotes the weaker coupling connecting the hybridized pair to the rest of the network.
Although this minimal model captures the scaling behavior that appears in numerical simulations, it does not {\it quantitatively} describe the results. To address this shortcoming, we introduce a more detailed model. 

We enrich the minimal three-site model by introducing an additional structure. In this extended model, \emph{three} energy scales associated with the tunnel couplings are introduced: (i) $\epsilon_{\mathrm{bath}}$, a characteristic coupling within the large network (bath); (ii) $J$, the strong tunneling amplitude between the excitation site and the ``gateway'' site. These two sites hybridize and effectively localize the excitation when $J$ is large; and (iii) $\epsilon_{\mathrm{link}}$, the characteristic coupling between the gateway site and the rest of the bath. 

The distinction between $\epsilon_{\mathrm{link}}$ and
$\epsilon_{\mathrm{bath}}$ is essential for analyzing the relaxation dynamics, as these two scales control different stages of equilibration: the initial escape from the strongly hybridized cluster and the subsequent spreading of excitation within the surrounding disordered environment.

These three energy scales are illustrated in Fig.~\ref{fig:link}. The hierarchy of energy scales, $J > \epsilon_{\mathrm{link}} > \epsilon_{\mathrm{bath}}$, gives rise to hierarchical relaxation dynamics. To investigate the emergence of the long time scale $J^2/(\Gamma \epsilon_{\mathrm{link}}^2)$, we artificially tune $J$ to take values above and below the range of $\epsilon_{\mathrm{link}}$.  
We now elaborate on the setup shown in Fig.~\ref{fig:link}. The system comprises a ensemble of $N=10$ spins randomly distributed within a circular region. Within this ensemble, we identify a subset of $N-1$ spins that form a weakly coupled subcluster, characterized by a typical coupling scale $\epsilon_{\mathrm{bath}}$, set by the spin concentration and dipolar geometry. These spins are indicated in blue in Fig.~\ref{fig:link}. Another site (colored in red) serves as the excitation site. 
So far, this corresponds to our standard setup. To enable a {\it controlled} investigation of hierarchical relaxation dynamics, we introduce an additional spin, referred to as the \emph{gateway site}. It is placed in the vicinity of the excitation site, with a tunable coupling $J$ between them. We vary $J$ by adjusting the distance between the excitation and gateway sites, allowing it to span the range $10^{-4} \le J \le 1$. 
In the regime of large $J$ and weak dephasing, the ``twin fortress'' effect will emerge, leading to long-lived localization of the excitation within the excitation–gateway pair.

The gateway site mediates the transfer of excitations to the rest of the network. We denote by $\epsilon_{\mathrm{link}}$ the characteristic coupling scale that governs the transfer of excitations from the excitation–gateway cluster into the rest of the network. This scale reflects the typical strength of the dipolar interactions connecting the excitation–gateway pair to the remaining spins, and thus depends on the full $N$-spin geometry.
The energy scale $\epsilon_{\mathrm{bath}}$ in contrast is the characteristic interaction scale governing redistribution of excitation within the residual weakly-coupled network, once population has escaped from the excitation subcluster. This scale is determined by the typical nearest-neighbor connectivity among the remaining $(N-1)$ spins.


\subsection{Results}

To demonstrate the impact of the energetic hierarchy beyond specific cases, we select 100 random configurations of $N=10$ spins, such as shown in Fig.~\ref{fig:link}.
For each realization, we extract the integrated survival measure $T_{ii}$ with $i$ as the excitation site, and analyze its statistical distribution.
Rather than averaging over realizations, we focus on the median relaxation time in order to suppress the influence of rare extreme configurations and capture the typical dynamical behavior.
The results for $T_{ii}$ as a function of $\Gamma$ for different values of $J$ are presented in Fig.~\ref{fig:Tii_vs_gamma}, and these are analyzed in detail in Fig. \ref{fig:combined_fits}.
Bootstrap uncertainties were calculated using 400 resamples. This allows us to demonstrate that the enriched, generalized three–energy-scale model yields quantitatively accurate predictions.




In Fig.~\ref{fig:combined_fits}, we analyze three key aspects of the problem, namely the dependence on $J$ of: 
(i) the survival-time measure $T_{ii}$ - as we show in Fig.~\ref{fig:combined_fits}(a), in the weak-dephasing regime and for large $J$, this scaling is approximately quadratic in $J$;
(ii) the value of the dephasing rate $\Gamma^{\ast}$ that minimizes $T_{ii}$ - Fig. ~\ref{fig:combined_fits}(b) shows that for large $J$, $\Gamma^{\ast}$ evolves linearly with $J$ toward larger values;
(iii) the corresponding minimal value of $T_{ii}$ - as shown in Fig.~\ref{fig:combined_fits}(c), for large $J$ this time grows linearly with $J$.

Beyond scaling relations, we find that the parameter choices $\epsilon_{\mathrm{bath}} \approx 2 \times 10^{-4}$ and $\epsilon_{\mathrm{link}} \approx 8 \times 10^{-3}$ provide a consistent description of the data. These values are in reasonable agreement with the case study presented in Fig.~\ref{fig:Jij}.

We discuss next in details the scaling results of Fig.~\ref{fig:combined_fits}.
In the weak dephasing regime, the strongly coupled dimer (excitation-gateway) hybridizes with strength $J$, while leakage to the larger subnetwork (characterized by the scale $\epsilon_{\mathrm{bath}}$) is off-resonant and dephasing-assisted. The effective relaxation rate to the large network scales as
$t_{\rm tr}^{-1} \sim \frac{\Gamma \epsilon_{\mathrm{link}}^2}{J^2}$,
leading to the survival-time measure
$T_{ii} \sim \frac{J^2}{\Gamma \epsilon_{\mathrm{link}}^2}$.
%
At fixed small $\Gamma$, we therefore expect $T_{ii} = \alpha J^2$, which is confirmed in Fig.~\ref{fig:combined_fits}(a). Substituting $\Gamma \sim 3 \times 10^{-5}$ and $\epsilon_{\mathrm{link}} \sim 8 \times 10^{-3}$ yields a prefactor $\alpha \approx 5 \times 10^{8}$, in good agreement with the simulations.

In the intermediate dephasing regime, $\epsilon < \Gamma < J$, two relaxation timescales, namely, the leakage from the dimer to the large network (bath) and the relaxation within this network, become comparable. Equating these timescales allows us to estimate the dephasing at  the minimum relaxation time,
\begin{align}
T_{\rm min} \sim \frac{\Gamma^*}{\epsilon_{\mathrm{bath}}^2}
\;\approx\;
\frac{J^2}{\Gamma^* \epsilon_{\mathrm{link}}^2}.
\end{align}
From this relation, we obtain
\begin{equation}
\Gamma^* \approx J \frac{\epsilon_{\mathrm{bath}}}{\epsilon_{\mathrm{link}}},
\end{equation}
the characteristic dephasing strength at which the minimum occurs.
In simulations we obtain $ \Gamma^{\ast} \simeq 0.025\,J$, see  Fig.~\ref{fig:combined_fits}(b), which agrees with  $\epsilon_{\mathrm{bath}}/\epsilon_{\mathrm{link}} =0.025$ using our chosen parameters.

Furthermore, substituting $\Gamma^{\ast}$ into $T(\Gamma) = \frac{J^2}{\Gamma \epsilon_{\mathrm{link}}^2}$ yields
\begin{align}
T_{\min}
\sim
\frac{J}{\epsilon_{\mathrm{bath}} \, \epsilon_{\mathrm{link}}},
\end{align}
predicting a linear scaling of the minimum time with $J$, consistent with Fig.~\ref{fig:combined_fits}(c). Using our parameter values, the slope falls within the expected order of magnitude, $1/(\epsilon_{\mathrm{bath}} \epsilon_{\mathrm{link}})\approx 6.3 \times 10^{5}$.

Thus, our minimal model, with a controlled excitation–gateway coupling, provides another major result of this paper. It demonstrates that a single strong bond within a fixed geometry is sufficient to shift the transport optimum and generate hierarchical relaxation dynamics. Importantly, this mechanism arises purely from geometric structure and does not rely on increasing the bath size.


\section{Ensemble Statistics of Finite Spin Baths at fixed density}
\label{sec:ensemble}

\begin{figure*}
    \centering
    \includegraphics[width=1\linewidth]{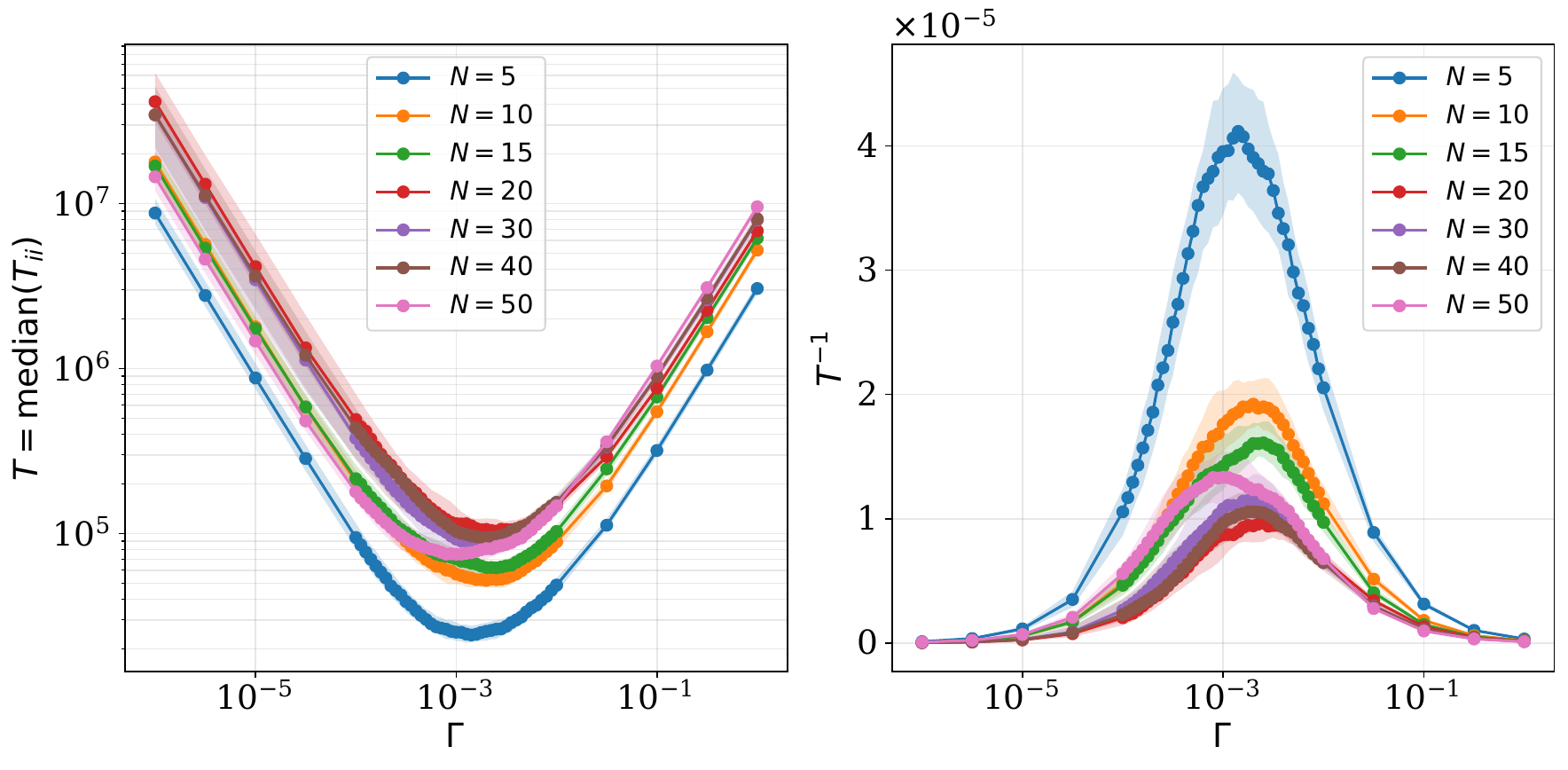}
    \caption{(left) Median survival time measure $T_{ii}$ and (right) inverse survival time measure $T_{ii}^{-1}$ as a function of $\Gamma$ for bath with $N=5$ sites to $60$. A global downward trend in $T_{ii}$ is observed as $N$ increases. 
    The radius is scaled with system size to maintain constant density, with $R=16$ nm for $N=10$ and $R_N = 16 (N/10)^{0.5}$ for general $N$. }
\label{fig:tii_vs_gamma_fixed_density}
\end{figure*}


    


\begin{figure*}
    \centering
    \includegraphics[width=1\linewidth]{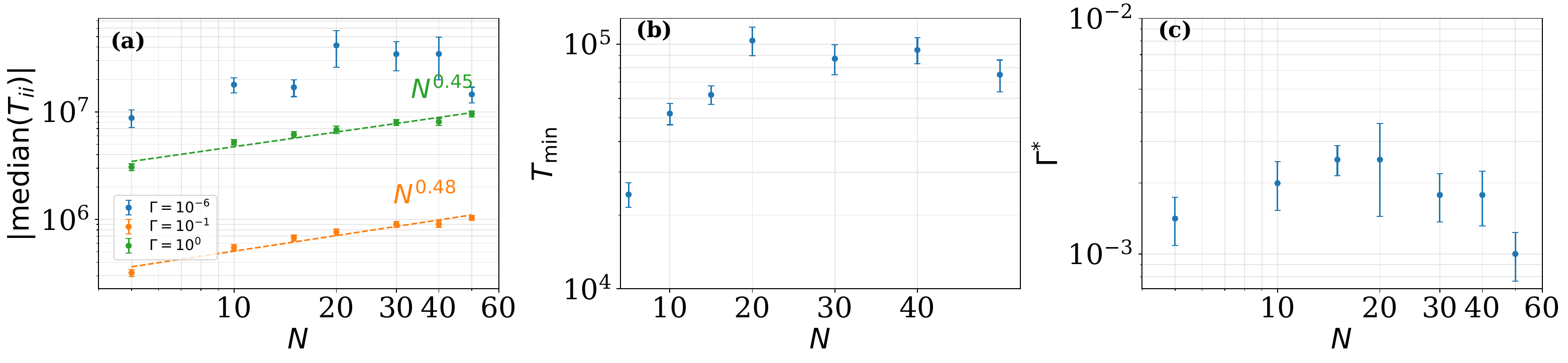}
\caption{System size scaling of $T_{ii}$.  
(a) Median $|T_{ii}|$ as a function of system size $N$ at a fixed dephasing strengths $\Gamma$, showing a power law scaling with $N^{\alpha}$. For the large dephasing regime, $\Gamma = 10^{-1}$ and $\Gamma = 10^{0}$, the scaling is approximately $N^{0.5}$. 
In contrast, for weak dephasing, $\Gamma = 10^{-6}$, the dependence on $N$ is very weak.
(b) Minimum value of the survival time, $T_{\min}$, presented as a function of $N$. An initial increase is observed, followed by a saturation around $N \sim 20$. 
(c) Optimal dephasing $\Gamma^*$, defined as the value of $\Gamma$ at which $T_{\min}$ is attained, plotted as a function of $N$. No strong dependence on system size is observed, consistent with the approximately invariant coupling statistics at fixed density.}
\label{fig:tii_scaling_fixed_density}
\end{figure*}

So far, we examined various geometries while keeping the system size fixed.
However, in realistic disordered spin environments, transport may occur in larger networks where geometric heterogeneity is amplified and multiple clusters and pathways coexist. This motivates a systematic investigation of larger ensembles, aimed at determining the robustness of hierarchical relaxation dynamics and identifying statistical features that persist across different disorder realizations.

To probe these effects, we study system-size scaling at fixed spin density, ensuring that larger systems represent extended realizations of the same underlying disordered structure. Consequently, the radius of the system is scaled as $R \propto \sqrt{N}$, maintaining spatial statistics and the coupling distribution as the bath grows. This construction enables us to isolate intrinsic size-dependent behavior and assess the robustness of hierarchical relaxation in larger networks.

To ensure comparable statistics across system sizes, we fix the total number of sampled sites to $\sim 2000$ by adjusting the number of disorder realizations as $\sim 2000/N$. Thus, smaller systems are averaged over more realizations, while larger systems are sampled with fewer, maintaining a constant statistical weight. For each realization, we compute the site-resolved survival-time measure $T_{ii}$ for all sites $i$ over a logarithmically spaced range of dephasing strengths $\Gamma$, enabling a systematic analysis of relaxation statistics across system sizes.

We begin by examining the survival-time measure across different system sizes. As shown in Fig.~\ref{fig:tii_vs_gamma_fixed_density}, the survival curves $T_{ii}(\Gamma)$ exhibit a well-defined minimum for all values of $N$. Notably, the position of this minimum remains approximately unchanged as the system size increases. This observation is further supported by Fig.~\ref{fig:tii_scaling_fixed_density}(c), where no systematic dependence of $\Gamma^*$ (the dephasing strength at which $T_{ii}$ is minimized) on $N$ is observed.

This size-independence indicates that the optimal dephasing scale is governed by local coupling statistics rather than global system size. In other words, the competition between coherent hybridization and dephasing-induced transfer, which sets $\Gamma^*$, is controlled by typical local energy scales that remain invariant at fixed density.

We next examine the system-size dependence of the median survival time at fixed dephasing strengths, shown in Fig.~\ref{fig:tii_scaling_fixed_density}(a). For weak dephasing, $\Gamma \approx 10^{-6}$, no clear dependence on $N$ is observed. This reflects the fact that, in this regime, the dynamics are governed by $1/\Gamma$ and by ratios of coupling strengths, which remain invariant under fixed-density scaling.
Consequently, relaxation is dominated by local cluster dynamics, leading to an approximately size-independent survival time.
In contrast, for stronger dephasing, $\Gamma = 10^{-1}$ and $\Gamma = 10^{0}$, the median survival measure increases approximately as $N^{1/2}$. Since the system's radius scales as $R \propto N^{1/2}$, this behavior suggests that in the strong dephasing regime the relevant relaxation scale grows with the linear size of the bath. 
Since $T_{ii}$ characterizes the equilibration timescale, it is natural to expect that larger systems generally require more time $\propto R$ to reach equilibrium. 

Next, we focus on the minimum value of the survival-time measure, $T_{\min}$. As shown in Fig.~\ref{fig:tii_scaling_fixed_density}(b), $T_{\min}$ initially increases with the size of the system before saturating around $N \sim 20$. This saturation marks the onset of a large-bath regime, where further increases in $N$ do not significantly affect the relaxation timescale.
This behavior reflects the proliferation of parallel relaxation pathways: beyond a certain size, the excitation can escape through multiple channels, eliminating global bottlenecks.

In Fig.~\ref{fig:tii_scaling_fixed_density}(c), we find that the optimal dephasing strength $\Gamma^*$ remains approximately independent of system size at fixed density. This follows from the fact that local coupling strengths $J_{ij}$ are statistically invariant with $N$ (see Appendix~\ref{AppB}). As $\Gamma^*$ is determined by local energy scales, it remains insensitive to the size of the system.

In Appendix~\ref{AppC}, we repeat this analysis while fixing the system size and increasing the spin density, thereby isolating the effect of enhanced local coupling strengths. In contrast to Fig.~\ref{fig:tii_vs_gamma_fixed_density}, increasing the density shifts the optimal dephasing $\Gamma^*$ to higher values and reduces the survival time, consistent with the strengthening of tunnel couplings throughout the network.


\section{Summary and Outlook}
\label{sec:summary}

In this work, we investigated quantum transport and relaxation dynamics in spatially disordered spin networks subject to local dephasing. While our approach was motivated by experiments with NV centers in diamond, our analysis is relevant for an extremely broad range of quantum transport scenarios, and provides insights into open questions associated with optimal coherent transport, and the interplay of interactions and disorder. We focused on two dimensional tight binding models with dipolar couplings and local dephasing.
Our main result can be summarized as follows: 

(R1) {\bf Hierarchical relaxation in disordered systems.} 
We identified a generic mechanism by which geometry generates emergent slow dynamics. Geometric disorder naturally induces a hierarchy of tunneling amplitudes, which, in turn, gives rise to multiple well-separated relaxation timescales. 
These timescales manifest as distinct dynamical stages in the relaxation process, including rapid coherence decay, equilibration within strongly coupled clusters, and much slower global relaxation governed by weak inter-cluster connectivity.
 
To understand the microscopic origin of the slowest relaxation process, we introduced a minimal three site model consisting of a strongly hybridized pair weakly coupled to a remote site. Strong internal coupling  generates an effective energy mismatch, which suppresses population transfer in the weak dephasing regime. 
This mechanism produces a parametrically enhanced relaxation time scaling as
$t \sim \frac{J^2}{\epsilon^2 \Gamma}$, which can exceed the homogeneous bath relaxation scale by several orders of magnitude. 

Using a combination of dynamical measures, steady-state transport analysis, and ensemble averaging, we showed that these hierarchical relaxation processes were robust, persisting across disorder realizations. We further demonstrated that environmental dephasing not only controls the overall relaxation rate but also governs the selection and competition between distinct transport pathways, leading to optimal transport regimes and, in certain configurations, multiple transport optima.

More broadly, our results demonstrate that relaxation in finite dipolar spin baths is governed not by typical nearest neighbor diffusion but by geometric hierarchy. Rare strongly-coupled clusters generate slow spectral modes of the Liouvillian, which control global equilibration. In this sense, the relaxation dynamics contain information about the geometric structure of the bath itself.

Using the physical normalization introduced in Sec. \ref{sec: Physical Setup, Model and dynamical Measures}, where \(J_{\max}\sim 1\,{\rm GHz}\), one simulation time unit corresponds to approximately \(1\,{\rm ns}\). We focus in particular on the regime \(\Gamma\sim 10^{-4}-10^{-3}\), where the dephasing rate is comparable to the weak link couplings \(\epsilon\) in the configurations studied here. This regime is experimentally natural if the effective dephasing originates from spin dynamics not explicitly included in the coherent Hamiltonian, such as flip-flop processes of additional bath spins or magnetic noise from \(^{13}{\rm C}\). With this conversion, the typical relaxation in Fig. \ref{fig:geomA_dynamics}(c)-(d), around \(t\sim 10^4\), corresponds to approximately \(10\,\mu{\rm s}\), while the hierarchical bottleneck dynamics can extend to \(t\sim 10^6\), corresponding to approximately \(1\,{\rm ms}\). Thus, geometric hierarchy can prolong otherwise natural microsecond scale NV spin bath dynamics by roughly two orders of magnitude, producing experimentally relevant millisecond scale relaxation tails.

Our work offers several additional important  contributions.

(R2) {\bf Dynamical Measures.} We introduced and analyzed integrated timescale-related measures to characterize survival and transfer dynamics, $T_{ii}$ and $T_{ij}$, respectively. We expect these measures to be useful in other systems. 

(R3) {\bf Twin-peak phenomenon in environment-assisted transport.} By constructing the steady state and examining competing transport pathways, we showed that two-dimensional disordered systems can exhibit distinct optimal regimes for quantum transport, each optimized at different dephasing strengths. This provides a mechanistic explanation of, and a framework to engineer, the ``twin-peak'' phenomenon recently observed in quantum transport~\cite{Eriktwin23}.


The emergence of well-separated relaxation timescales, local cluster-level equilibration prior to global equilibration, and bottleneck-controlled transport bears conceptual similarity to glassy relaxation phenomena, where heterogeneous structures generate metastable dynamical regimes and broad relaxation spectra \cite{glass2}.

Our findings have direct implications for experimentally accessible solid-state spin platforms, such as NV centers in diamond, where disordered spin environments naturally give rise to strongly heterogeneous coupling networks. More broadly, the mechanisms identified here are expected to apply to a wide class of disordered quantum systems, including excitonic transport in molecular networks, nanoscale energy transfer, and quantum information processing architectures.

Several experimentally-motivated extensions of this work can be envisioned. In particular, extending the analysis to multiple excitations would enable a direct comparison between single-particle transport and many-body relaxation processes, such as those underlying the dynamics of $T_1$ and $T_2$ processes and would provide insight into how interactions and collective effects modify the transport mechanisms identified here.
Incorporating correlated noise and geometry-dependent fluctuations could further elucidate the role of environmental memory in hierarchical relaxation dynamics.  
Finally, extending the framework to larger networks and to multiple excitation centers may enable exploration of bath-mediated correlations and information transfer in solid-state spin platforms.





\begin{acknowledgments}
The work of M.L. is supported by the NSERC Canada Graduate Scholarship-Doctoral. 
D.S. acknowledges support from an NSERC Discovery Grant and an NSERC Alliance International Catalyst Grant.
R. N. acknowledges the 
International Visiting Graduate Student program (IVGS) at
the University of Toronto for facilitating his research visit to the University of Toronto.
B.M. acknowledges the support of an Ontario Graduate Scholarship Award.  The work of M.L. is supported by the NSERC Canada
Graduate Scholarship-Doctoral.

N.B. acknowledges support by the European Commission’s Horizon Europe Framework Programme under the Research and Innovation Action GA No. 101070546-MUQUABIS and ERC CoG Project QMAG (no. 101087113). N.B. also acknowledges financial support by the Ministry of Science and Technology, Israel, the Vatat quantum computing centers initiative, the innovation authority (project DiamondSemiIL no. 88708), and the ISF (Grants No. 1380/21 and No. 3597/21).

\vspace{10mm}
{\bf Data Availability}
The data that support the findings of this article are openly available at \cite{dataN}.


\end{acknowledgments}
%
\newpage

\appendix

\section{Dynamical measures}
\label{AppA}

In this Appendix, we discuss different measures for the dynamics, starting from the spectrum of the Liouvillian. We then
clarify the relation between integrated population measures, spectral properties of the Liouvillian, and the definition of relaxation and transfer times used throughout the paper.

\subsection{Liouvillian spectrum} 

\begin{figure}[t!]
\includegraphics[width=0.9\linewidth]{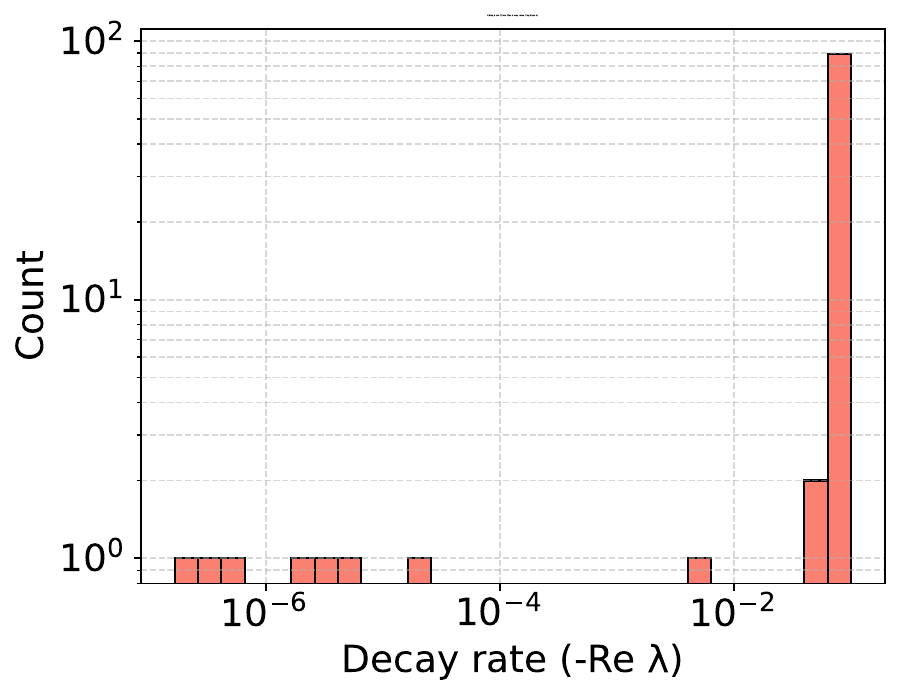}
\caption{Configuration A: Histogram of the eigenvalues of the Liovillian. Parameters are the same as in Fig. \ref{fig:geomA_dynamics} with $\Gamma$=0.1.
}
\label{fig:rates}
\end{figure}

\begin{figure}[t!]
\includegraphics[width=0.9\linewidth]{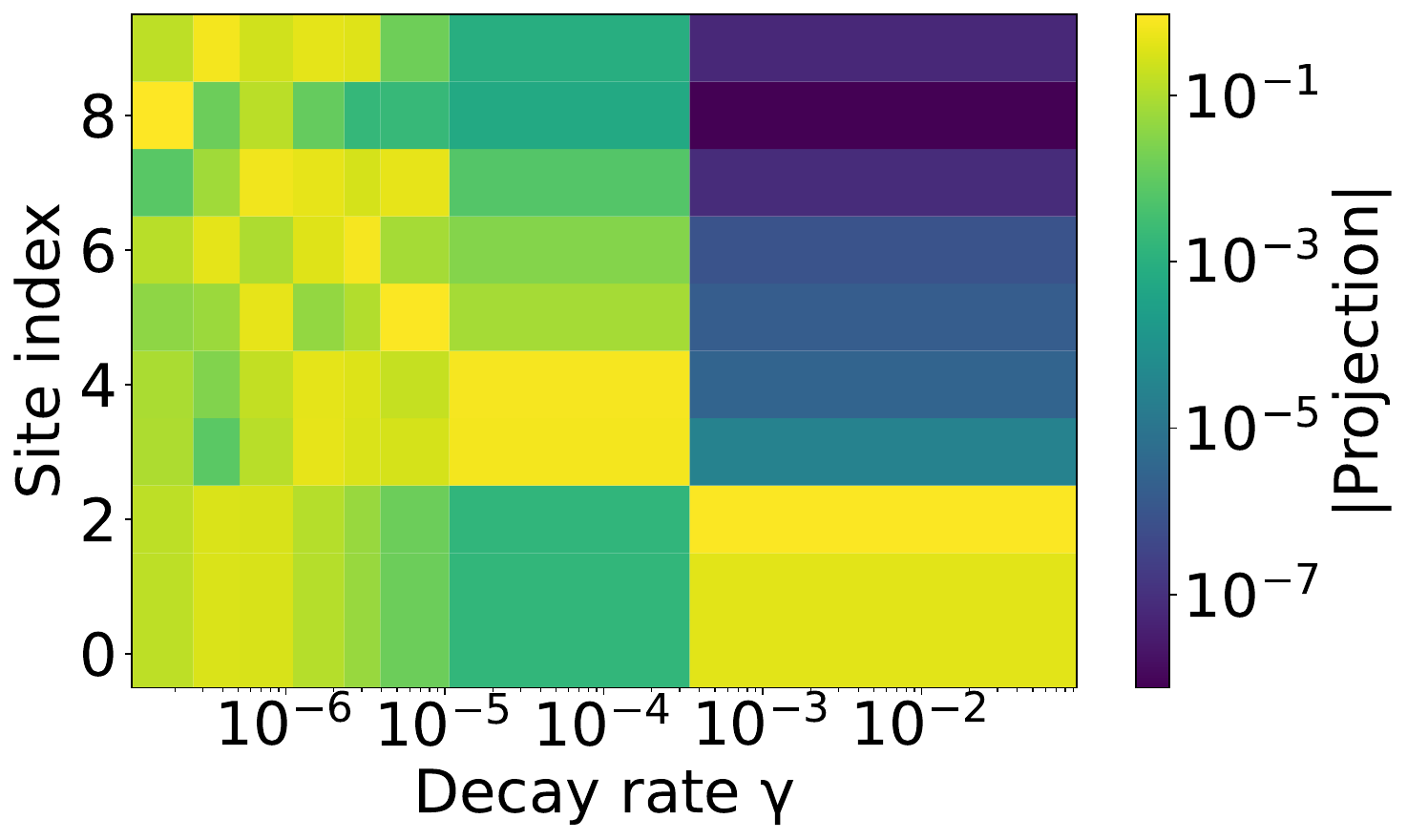}
\caption{Configuration A: Projection of the slow eigenvectors of the Liovillian on the different sites (0 to 9).
Here the first 89 values were ignored,along with the steady state vector .
}
\label{fig:ratesM}
\end{figure}

Dissipative dynamics can be characterized by analyzing the eigenvalues and eigenstates of the Liouvillian, Eq. (\ref{lindpd}). The decay rate, denoted by $\gamma$, is given by the negative real part of each eigenvalue, while the imaginary part determines the corresponding oscillation frequency.
We present our results for configuration A in Figs.~\ref{fig:rates}-\ref{fig:ratesM}. 

We begin by examining the real parts of the Liouvillian eigenvalues in Fig.~\ref{fig:rates}.
In the strong dephasing regime ($\Gamma = 0.1$), the majority of eigenvalues (89 out of 100)  exhibit rapid decay with rates $\gamma \approx 0.1$. These modes correspond to fast dynamics occurring on timescales of order $1/\Gamma$.
In contrast, a smaller subset of 10 eigenvalues displays significantly reduced decay rates. These slower modes are associated with inter-cluster transitions, thus they govern the long-time equilibration dynamics of the system.
Note that the zero eigenvalue of the Liovillian is excluded from this figure.

To better understand what the slower eigenvalues are, in Fig. \ref{fig:ratesM} we inspect the corresponding eigenvectors, projected onto the different sites populations, 0 to 9. 
The corresponding population dynamics in Fig. \ref{fig:geomA_dynamics}(e)-(f)
assists in realizing the processes.

The three modes with decay rates $\gamma \approx 0.01$--$0.1$ exhibit a large weight on sites $0,1,2$, which are spatially proximate (see the geometry in Fig.~\ref{fig:geom1}). These relatively large rates correspond to rapid equilibration within this triplet, with $\gamma \approx J^2/\Gamma$, where $J \sim 0.1$ and $\Gamma \sim 0.1$.
Next, the eigenvector associated with $\gamma \approx 10^{-5}$ has significant support at sites $3$ and $4$, indicating local equilibration within this pair. The corresponding rate again follows $\gamma \approx J^2/\Gamma$, here with a much smaller coupling $J \sim 10^{-3}$.
Finally, the remaining six small eigenvalues describe slow equilibration processes between different clusters. These rates are governed by weak inter-cluster couplings $\epsilon \approx 10^{-5}$--$10^{-4}$, leading to $\gamma \approx \epsilon^2/\Gamma$.


\subsection{MFPT and residence times}

We study the dynamics using the following measures: population dynamics from a certain initial condition, purity of the state 
${\rm Tr}[\rho(t)^2]$, and 
and measures for 
survival and transfer times, which we now define by modifying the traditional definitions of the mean first passage time.

Consider an $N+2$ site system; the network itself includes $N+1$ sites where we (arbitrarily) assign the initial condition to site $0$, and wish to study population transfer to site $N$. From site $N$, population leaves irreversibly to site $T$ (trap). 
Let us define the mean first passage time (MFPT) to arrive at site $T$, $\tau_M$ as follows
\bea
\tau_M=\int_0^{\infty} d\tau \tau \dot \rho_{TT}(\tau).
\eea
What is the interpretation of this expression?
$\rho_{TT}(t)$ is the population accumulated at site $T$ at time $t$. Its derivative is the instantaneous population change. Suppose we discreteize the MFPT into short time intervals: The average arrival time is given by the net
number of particles arriving at site $T$ within a unit time interval, multiplied by the time interval. We sum then over the full range of time, from initial conditions to equilibrium.

We now note the normalization condition
$1=\sum_{j=0}^{N}\rho_{jj}(t) +\rho_{TT}$, thus
$\dot \rho_{TT}=-\sum_{j=0}^{N}\dot\rho_{jj}$.
Plugging this expression to the MFPT definition, we get
\bea
\tau_M=\sum_{j=0}^{N}\int_0^{\infty} d\tau 
\tau \dot \rho_{jj}(\tau)  
\label{eq:tauM}
\eea
We integrate by parts; here we make use of the  assumption that in the long time limit, all the population sits in the trap site thus $\rho_{jj\neq T}(t\to \infty)=0$.
As a result, we get
\bea
\tau_M&=&\sum_{j=0}^N \int_0^\infty \rho_{jj}(\tau) d\tau
\nonumber\\
& =&  \sum_j\tau_j,
\eea
Where we defined $\tau_j=\int_0^{\infty} d\tau \rho_{jj}(\tau)$.
Now, in a vectorized notation, $\rho(t)=e^{{\hat{\mathcal L}}t}\rho(0)$
thus
\bea
\tau_j=[\hat {\mathcal {L}}^{-1}\hat \rho(0)]_{jj}
\eea
which is the residence time at each site.

Note that this approach requires us to add decay from site $N$, so we do need to change the Liouvillian and add the $\gamma_l$ in the same fashion as done in the flux calculation! The critical assumption was that populations in the physical system go to zero at long time.

\subsection{Integrated relaxation and transfer measures}

We consider open system dynamics generated by a Liouvillian superoperator $\hat{\mathcal L}$,
\begin{equation}
\dot{\hat \rho}(t)=\hat {\mathcal L}\hat \rho(t), \qquad \hat \rho(t)=e^{\hat{\mathcal L} t}\hat \rho(0).
\end{equation}
Assuming that population in the physical subspace decays at long times, as ensured for example by introducing an absorbing trap,
\begin{equation}
\langle i |\rho(t\to\infty) |i \rangle=0, \,\,\,\, \forall i
\end{equation}
the time integrated density matrix is well defined and satisfies
\begin{equation}
\int_0^\infty \hat \rho(t)\,dt=\hat{\mathcal L}^{-1}\hat \rho(0).
\end{equation}
Projecting onto a site population operator $\rho_{jj}=\langle j| \rho| j \rangle$ yields
\begin{equation}
\int_0^\infty \rho_{jj}(t)\,dt=[\hat{\mathcal L}^{-1}\hat \rho(0)]_{jj},
\end{equation}
which we interpret as the residence time of the excitation at site $j$.

This expression admits an equivalent spectral representation.
Let $\{\lambda_k,|R_k\rangle,\langle L_k|\}$ denote the eigenvalues and biorthogonal eigenvectors of $\hat{\mathcal L}$,
\begin{equation}
\hat{\mathcal L}|R_k\rangle=\lambda_k|R_k\rangle, \qquad
\langle L_k|\hat{\mathcal L}=\lambda_k\langle L_k|.
\end{equation}
Formally,
\begin{equation}
\hat{\mathcal L}^{-1}=\sum_k\frac{1}{\lambda_k}|R_k\rangle\langle L_k|.
\end{equation}
Applying this representation to an initial site population $\rho_i\equiv|i\rangle\langle i|$ and projecting onto site $j$ yields
\begin{equation}
T_{ij}=[\hat{\mathcal L}^{-1}\rho_i]_{jj}
= \sum_k \frac{\langle \rho_j|R_k\rangle\langle L_k|\rho_i\rangle}{\lambda_k}.
\end{equation}

In the absence of an absorbing mechanism, the Liouvillian possesses a zero eigenvalue associated with the steady state, and its inverse does not exist.
In practice, one may exclude this mode and define a pseudo inverse acting on the decaying subspace,
\begin{equation}
\hat{\mathcal L}^{+}=\sum_{\lambda_k\neq 0}\frac{1}{\lambda_k}|R_k\rangle\langle L_k|.
\end{equation}
This procedure regularizes time integrated observables and yields finite values for $T_{ij}$.
However, it is important to emphasize that this construction is mathematical in nature and does not by itself define a mean first passage time (MFPT).

Indeed, in the absence of a trap, the population at a physical site may increase and decrease over time, and $\dot\rho_{jj}(t)$ may change sign.
As a result, quantities such as
\begin{equation}
\int_0^\infty t\,\dot\rho_{jj}(t)\,dt
\end{equation}
are not guaranteed to be positive and generally do not admit a first passage interpretation.
A proper MFPT is obtained only when an absorbing trap is introduced, ensuring irreversible population flow, removal of the steady state from the physical subspace, and strict positivity of the resulting timescale.
Throughout this work, trapping based MFPTs are therefore used to define relaxation times, while integrated populations and the associated $T_{ij}$ matrix are employed as complementary diagnostics of relaxation and transport.

Finally, although the Liouvillian $\hat{\mathcal L}$ is generally non Hermitian and may possess complex eigenvalues and eigenvectors, the quantities $T_{ij}$ are real.
For real Liouvillian matrices, eigenvalues and eigenvectors appear in complex conjugate pairs, and each complex contribution to the spectral sum is 
accompanied by its conjugate,
\bea
&& 
\frac{\langle \rho_j|R_k\rangle\langle L_k|\rho_i\rangle}{\lambda_k}
+
\frac{\langle \rho_j|R_k\rangle^*\langle L_k|\rho_i\rangle^*}{\lambda_k^*}
\nonumber\\
& &=
2\,\mathrm{Re}\!\left(\frac{\langle \rho_j|R_k\rangle\langle L_k|\rho_i\rangle}{\lambda_k}\right),
\eea
ensuring that $T_{ij}\in\mathbb R$.
This result holds for both diagonal elements ($i=j$), corresponding to residence or relaxation times, and non-diagonal elements ($i\neq j$), which quantify integrated transport between sites.

\section{Local versus global geometric statistics}
\label{AppB}

As the system size increases at fixed density, we examine the corresponding changes in the interaction network. While the local environment experienced by a given site remains statistically invariant, global properties such as the extremal values of the coupling strengths $J$ vary with system size.

%
%

\subsection{Pairwise coupling distribution}

\begin{figure}[htbp]
    \centering
    \includegraphics[width=0.9\linewidth]{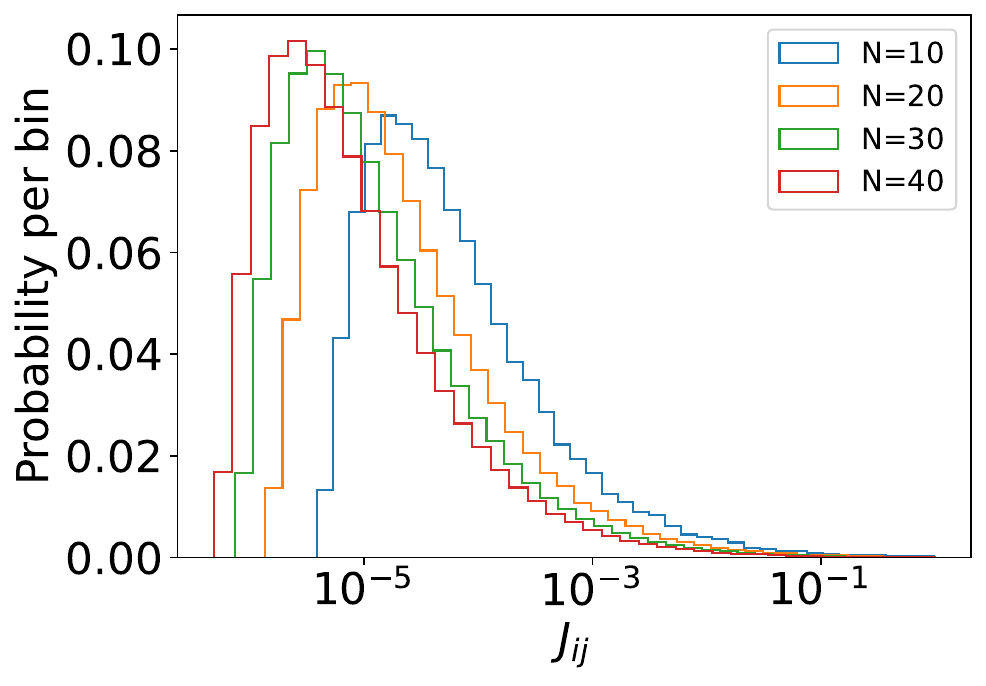}
    \caption{
Histogram of all pairwise couplings $J_{ij}$ upon increasing bath size $N$ at a fixed density. The distribution acquires additional weight at small $J_{ij}$ due to the inclusion of more distant pairs.  This reflects geometric enlargement of the system rather than a modification of the local environment experienced by a typical site. Histograms were generated by constructing 1000 configurations for each value of $N$. }
\label{fig:J_vals_hist}
\end{figure}

We first examine in Fig. \ref{fig:J_vals_hist} the histogram of all pairwise couplings, $J_{ij}$ across the bath.
As $N$ increases at a fixed density, the distribution develops an additional weight at small $J_{ij}$ values. This is expected: Enlarging the system adds more distant pairs, and since the interaction decays as a power law with distance, these additional pairs predominantly contribute weak couplings.
This behavior reflects the growth of the system size and the inclusion of larger radii. It does not by itself imply a modification of the local environment for a typical site.

\subsection{Local observables}

\begin{figure}
    \centering
    \includegraphics[width=1\linewidth]{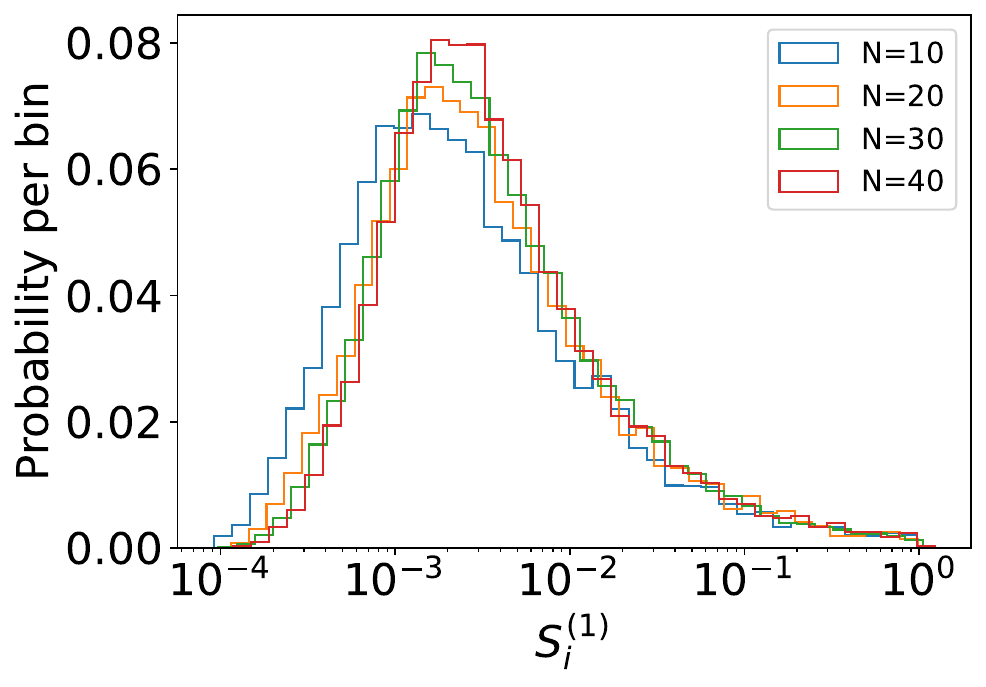}
    \caption{Histogram of the total coupling strength experienced by site $i$ $S_i^{(1)} = \sum_{j \ne i} J_{ij}$}
    \label{fig:s1_hist}
\end{figure}

\begin{figure}
    \centering
    \includegraphics[width=1\linewidth]{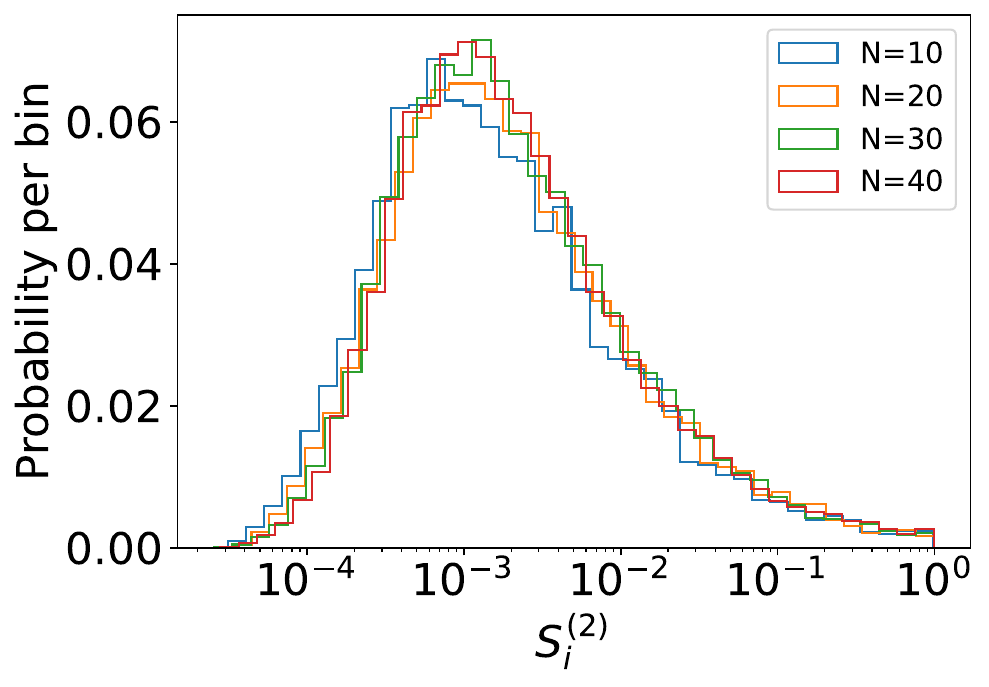}
    \caption{Histogram of quadratic accumulation of couplings for site $i$, $S_i^{(2)} = \sqrt{\sum_{j \ne i} |J_{ij}|^2}$}
    \label{fig:s2_hist}
\end{figure}

\begin{figure}[htbp]
    \centering
    \includegraphics[width=0.9\linewidth]{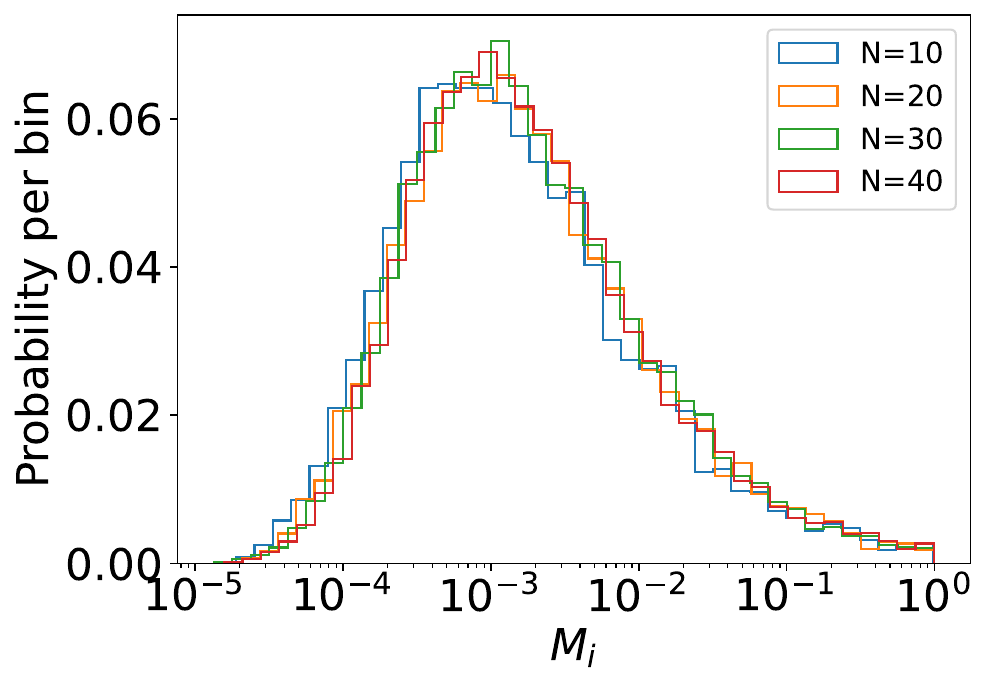}
    \caption{
Histograms of strongest local coupling, $M_i = \max_{j \ne i} |J_{ij}|$, experienced by individual sites. 
Apart from minor finite size effects at the smallest $N$, the distributions for $S_i^{(1)}$, $S_i^{(2)}$ and $M_i$ collapse across system sizes. This demonstrates that the local geometric environment of a typical site is statistically invariant when $N$ increases at fixed density.
}
\label{fig:Mi_hist}
\end{figure}

To focus on local effects, we now analyze site resolved observables defined ``row-wise" from the Hamiltonian matrix. We define the following observables: 
\begin{equation}
S_i^{(1)} = \sum_{j \ne i} |J_{ij}|,
\,\,\,
S_i^{(2)} = \sqrt{\sum_{j \ne i} |J_{ij}|^2},
\,\,\,
M_i = \max_{j \ne i} |J_{ij}|.
\end{equation}
Here, $S_i^{(1)}$ measures the total coupling strength experienced by site $i$; $S_i^{(2)}$ captures the quadratic accumulation of couplings for site $i$; 
$M_i$ identifies the strongest neighbor of site $i$.
Histograms for $S_i^{(1)}$, $S_i^{(2)}$, and $M_i$
are shown in Figs.~\ref{fig:s1_hist}, \ref{fig:s2_hist} and \ref{fig:Mi_hist}, respectively.
Strikingly, these distributions exhibit almost no systematic shift with increasing $N$, this effect is particularly notable for $M_i$. Apart from minor finite size effects at the smallest bath size, these histograms collapse nearly perfectly for larger $N$.  
This demonstrates that the local coupling environment of a typical site {\it remains statistically invariant} when $N$ is increased at a fixed density.



\subsection{Global extreme statistics}

\begin{figure}[htbp]
    \centering
\includegraphics[width=0.9\linewidth]{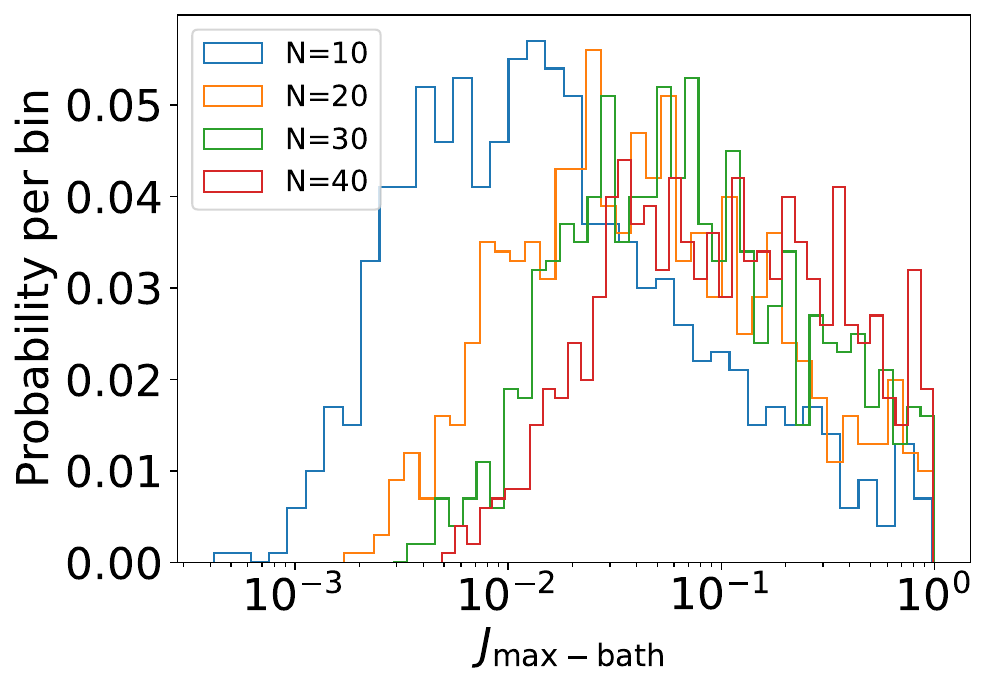}
    \caption{
Histogram of the bath maximum coupling $J_{\max}^{(\mathrm{bath})}$. 
In contrast to local observables, the distribution shifts systematically toward larger values as $N$ increases. 
This behavior reflects extreme value statistics: since the number of pairwise couplings scales as $N^2$, the probability of realizing a rare strongly coupled pair grows with system size. 
}
    \label{fig:Jmax_hist}
\end{figure}

We define the maximum coupling of the $N$-spin bath as
\begin{equation}
J_{\max}^{(\mathrm{bath})} = \max_{i,j} |J_{ij}|.
\end{equation}
In contrast to local observables, the histogram of $J_{\max}^{(\mathrm{bath})}$ systematically shifts toward larger values as $N$ increases; see Fig. \ref{fig:Jmax_hist}.
This behavior is expected since the total number of pairwise couplings scales as $N^2$, and therefore the probability of finding at least one unusually close pair increases with the size of the system. The strengthening of $J_{\max}^{(\mathrm{bath})}$ is thus a direct manifestation of extreme value statistics.
Crucially, this shift is not accompanied by any comparable modification of local statistics, as we demonstrated in Figs. \ref{fig:s1_hist}-\ref{fig:Mi_hist}. 
The size dependent restructuring of the bath therefore originates from global, rather than local, geometric effects.

\subsection{Physical interpretation}

The combined analysis of local and global observables leads to clear conclusions:
(i) Local geometric statistics are controlled by density and remain invariant with $N$. (ii) Global extreme couplings scale with system size.

\section{Ensemble statistics of finite spin baths: Fixed radius}
\label{AppC}

\begin{figure*}[htbp]
\centering
\includegraphics[width=0.8\textwidth]{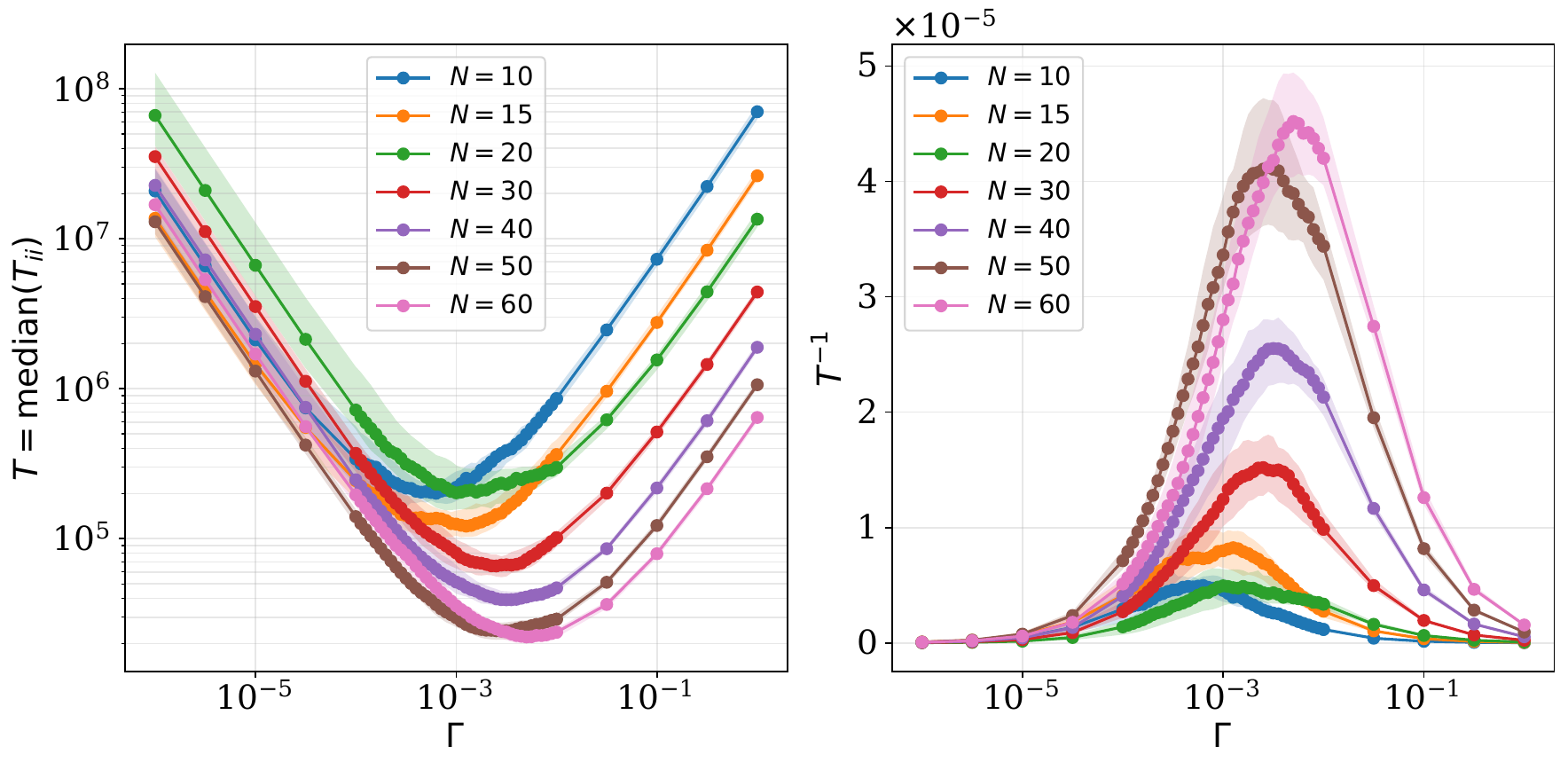}
\caption{
(left) Median survival measure $T_{ii}$ and (right) inverse survival measure $T_{ii}^{-1}$ as a function of $\Gamma$ for bath sizes $N=10$ to $60$ at a fixed radius $R$ and growing density. 
A global downward trend in $T_{ii}$ is observed as $N$ increases, indicating improved connectivity and reduced geometric isolation. 
At the same time, the location of the minimum shifts systematically toward larger $\Gamma$. 
Error bars represent bootstrap uncertainties.}
\label{fig:ch5_Tii_stats}
\end{figure*}

\begin{figure*}
    \centering
\includegraphics[width=1\linewidth]{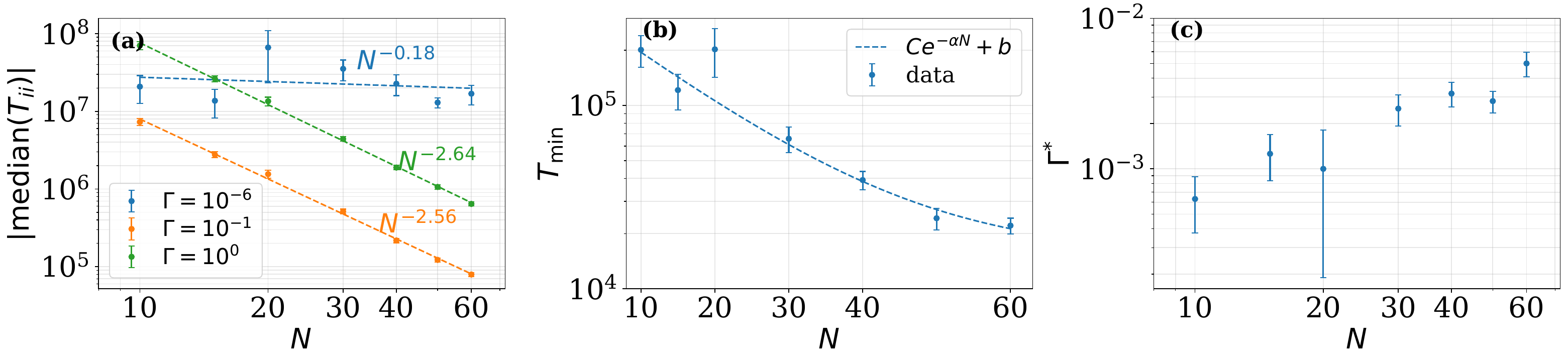}
\caption{
System size scaling with of survival time measure with increasing density.
(a) Median $|T_{ii}|$ as a function of system size $N$ at a fixed dephasing strength $\Gamma$, showing power law scaling with $N^{\alpha}$. For the large dephasing regimes, $\Gamma = 10^{-1}$ and $\Gamma = 10^{0}$, the scaling is approximately $N^{-2.6}$.
In contrast, for weak dephasing, $\Gamma = 10^{-6}$, the dependence on $N$ is very weak, suggesting almost no clear $N$ dependence.
(b) Minimum value $T_{\min}$ vs.\ $N$, fit to an exponential decay $C e^{-\alpha N} + b$, with $C = (3.6 \pm 0.9)\times 10^{5}$, $\alpha = (6.8 \pm 1.3)\times 10^{-2}$, and $b = (1.5 \pm 0.5)\times 10^{1}$.
(c) $\Gamma^*$, the value of $\Gamma$ at which $T_{\min}$ is attained, plotted as a function of $N$. A rightward shift of $\Gamma^*$ is observed as $N$ increases.
}
\label{fig:tii_scaling}
\end{figure*}

The analysis presented in Figs.~\ref{fig:ch5_Tii_stats}-\ref{fig:tii_scaling} correspond to a fixed radius setup, in which the spin density {\it increases} with system size.

For bath sizes $N=10$ to $60$, approximately fifty independent realizations were generated at each value of $\Gamma$. 
For each realization, the full set of site resolved survival measures $T_{ii}$ was calculated over fifty logarithmically spaced dephasing strengths $\Gamma$, with enhanced sampling between $10^{-4}$ and $10^{-2}$.
Since the distribution of $T_{ii}$ is strongly skewed due to rare configurations containing unusually strong local couplings, the median was chosen as the representative statistic rather than the mean.  
Uncertainties were estimated using bootstrap resampling across bath realizations.
The total number of sampled sites ranged from roughly five hundred for $N=10$ up to three thousand for $N=60$.

A clear trend emerges in Fig. \ref{fig:ch5_Tii_stats}. The typical survival time decreases as $N$ grows.  
As the spin density increases with $N$ at fixed radius, the probability of poorly connected edge sites is reduced and rare protective configurations are suppressed.
The position of the minimum $\Gamma^*$ shifts monotonically toward larger dephasing strengths as $N$ increases.  
This behavior mirrors the single configuration analysis, where stronger local couplings displace the optimal noise scale. The increasing spin density enhances the likelihood of stronger couplings, which in turn requires larger $\Gamma$ to suppress coherent bottlenecks and enable efficient transport.

In Fig. \ref{fig:tii_scaling} we further show that 
in the strong dephasing limit, 
the following scaling for the survival time measure develops, $T_{ii}\propto N^{-2.5}$. Explaining this trend requires further simulations and analysis. 
In contrast, the weak dephasing regime shows no clear dependence on concentration. This is interesting, but it is physically reasonable, since the relevant timescales in this regime are \(1/\Gamma\) for a homogeneous bath and \((J^2/\epsilon^2)/\Gamma\) for the hierarchical case. Under a uniform increase in density, both \(J\) and \(\epsilon\) increase by the same factor, so the ratio \(J^2/\epsilon^2\) remains unchanged. As a result, the characteristic weak dephasing timescales are expected to remain approximately constant.

The minimum survival time decreases approximately exponentially with density. This behavior indicates that protective substructures become statistically suppressed as the bath grows.  

The monotonic shift of $\Gamma^*$ with density constitutes another statistical feature of the system. 
As the density grows, the typical coupling strengths increase, leading to a systematic shift of the optimal dephasing scale.


\bibliography{ref} 

\end{document}